\documentclass[3p,twocolumn]{elsarticle}
\usepackage{lineno,hyperref}
\modulolinenumbers[5]
\usepackage{subcaption}
\usepackage{xcolor,colortbl}
\usepackage{graphicx}
\usepackage{hyperref}

\begin{document}

\definecolor{Gray}{rgb}{0.8,0.8,0.8}
\newcolumntype{g}{>{\columncolor{Gray}}c}

%%%%%%%%%%%%%%%%%%%%%%%%%%%%%%%%%%%%%%%%%%%%%%%%%%%%%%%%%%%%
\begin{frontmatter}

%\title{Deep neural networks heatmaps of Alzheimer's Disease classification using neuroimaging data capture meta-analysis patterns}

%\title{CNN Heatmaps Capture Alzheimer's Disease Effects Reported in Voxel-based Morphometry Studies}

\title{Deep neural network heatmaps capture Alzheimer's disease patterns reported in a large meta-analysis of neuroimaging studies}

\author[biggs]{Di Wang}
\author[biggs]{Nicolas Honnorat}
\author [rii]{Peter T. Fox}
\author [berlin]{Kerstin Ritter} 
\author [juelich,isn]{Simon B. Eickhoff}
\author [biggs]{Sudha Seshadri}
\author{the Alzheimer's Disease Neuroimaging Initiative \corref{adni}}
\author[biggs,rii]{Mohamad Habes \corref{corresponding}}
\cortext[corresponding]{Corresponding author}
\ead{habes@uthscsa.edu}
\address[biggs]{Neuroimage Analytics Laboratory and Biggs Institute Neuroimaging Core, Glenn Biggs Institute for Neurodegenerative Disorders, University of Texas Health Science Center at San Antonio, San Antonio, Texas, USA}
\address[rii]{Biomedical Image Analytics Division, Research Imaging Institute, University of Texas Health Science Center at San Antonio, San Antonio, Texas, USA}
\address[berlin]{Charite – University Medicine Berlin, Humboldt-University Berlin, Berlin, Germany}
\address[juelich]{Institute of Neuroscience and Medicine, Brain \& Behaviour (INM-7), Research Centre Jülich, Jülich, Germany}
\address[isn]{Institute of Systems Neuroscience, Medical Faculty, Heinrich-Heine University Düsseldorf, Germany}
\cortext[adni]{
The brain scans used in the present work were obtained from the Alzheimer's Disease Neuroimaging Initiative (ADNI) database (adni.loni.usc.edu) but ADNI investigators did not participate in the design, the analysis, or the writing of the manuscript. A complete listing of ADNI investigators can be found at http://adni.loni.usc.edu/wp-content/uploads/how\_to\_apply/ADNI\_Acknowledgement\_List.pdf.}

\begin{abstract}
Deep neural networks currently provide the most advanced and accurate machine learning models to distinguish between structural MRI scans of subjects with Alzheimer's disease and healthy controls. Unfortunately, the subtle brain alterations captured by these models are difficult to interpret because of the complexity of these multi-layer and non-linear models. Several heatmap methods have been proposed to address this issue and analyze the imaging patterns extracted from the deep neural networks, but no quantitative comparison between these methods has been carried out so far. In this work, we explore these questions by deriving heatmaps from Convolutional Neural Networks (CNN) trained using T1 MRI scans of the ADNI data set, and by comparing these heatmaps with brain maps corresponding to Support Vector Machines (SVM) coefficients. Three prominent heatmap methods are studied: Layer-wise Relevance Propagation (LRP), Integrated Gradients (IG), and Guided Grad-CAM (GGC). Contrary to prior studies where the quality of heatmaps was visually or qualitatively assessed, we obtained precise quantitative measures by computing overlap with a ground-truth map from a large meta-analysis that combined 77 voxel-based morphometry (VBM) studies independently from ADNI. Our results indicate that all three heatmap methods were able to capture brain regions covering the meta-analysis map and achieved better results than SVM coefficients. Among them, IG produced the heatmaps with the best overlap with the independent meta-analysis.
\end{abstract}

\begin{keyword}
Deep Learning \sep Alzheimer's disease \sep MRI \sep Explainable AI \sep Neuroimaging 
\end{keyword}
\end{frontmatter}
% \linenumbers

%%%%%%%%%%%%%%%%%%%%%%%%%%%%%%%%%%%%%%%%%%%%%%%%%%%%%%%%%%%%%%%%%%%%%%%%%%%%%

\section{Introduction}

\begin{figure*}
\includegraphics[width=1.0\linewidth]{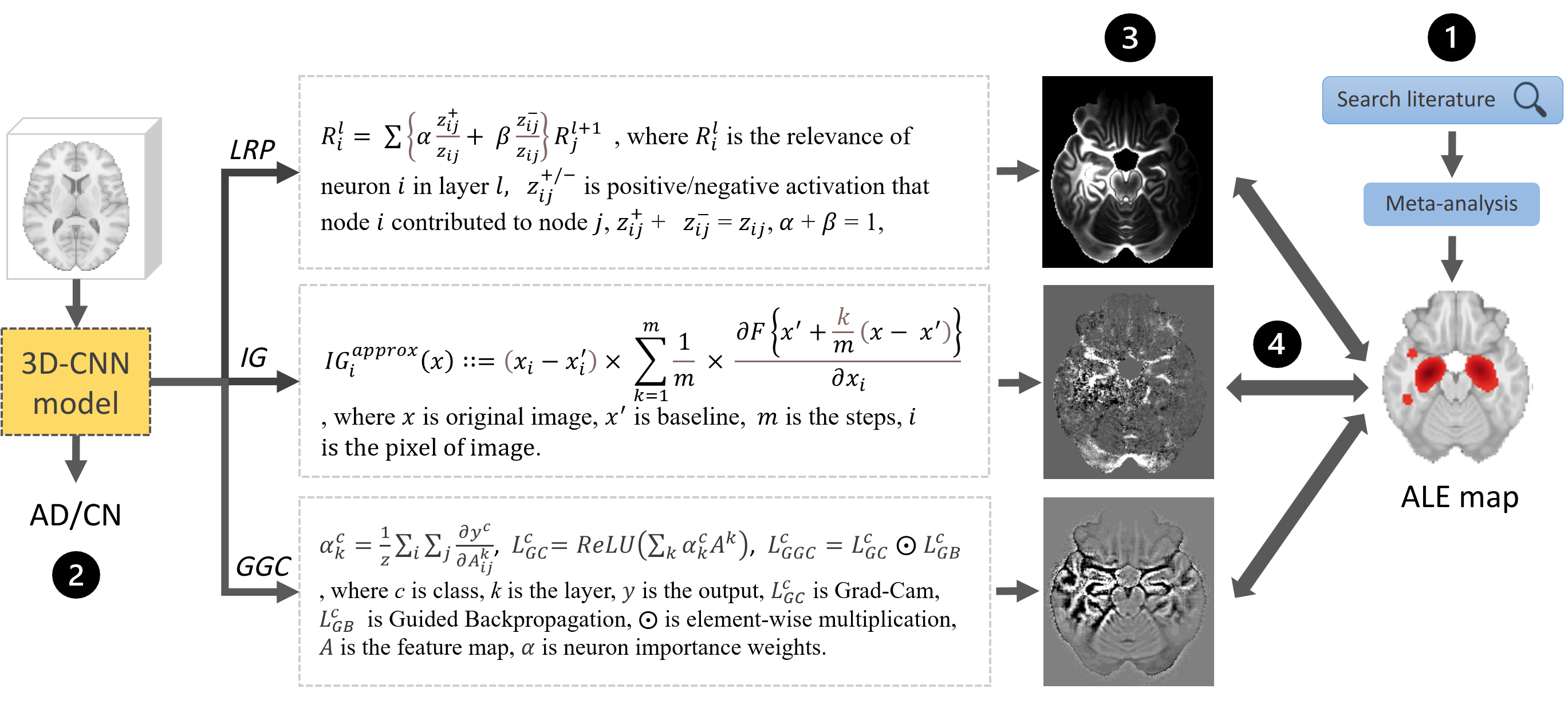}
\caption{\label{overview} An overview of the present study. (1) A VBM meta-analysis was conducted to derive an ALE map summarizing AD effects on the brain visible in T1-weighted MRI scans. (2) 3D CNNs were trained to classify AD and CN ADNI T1-weighted MRI scans. (3) Three heatmap methods were applied to the CNN model with the highest cross-validation accuracy. (4) The heatmaps were compared with the meta-analysis map.}
\end{figure*}

Alzheimer's disease (AD) is the most common brain dementia \cite{dementia_rush}. In 2020, 5.8 million Americans age 65 and older were living with AD, and this number is expected to reach 13.8 million by 2050 \cite{AD_usa}. Considerable efforts have been made to tackle the challenges raised by this issue and, in particular, research early neuroimaging biomarkers and prognosis tools \cite{AD_review, habes2016white, habes2016BA, Li2019DL}. The most recent Deep Learning frameworks were involved in these efforts and showed promising achievements in AD classification. For instance, an accuracy of 87.15\% was reported for a 3D CNN in a recent work, for the classification of T1-weighted MRI scans \cite{AD_CLF1}. A multi-instance 3D convolutional neural network reached an accuracy of 91.09\% for a similar task \cite{AD_CLF2}. Unfortunately, these machine learning frameworks rely on complex architectures which makes it difficult to understand what neurological changes are modeled by the deep networks as typical dementia signatures, markers of disease progression, and clues for differential diagnosis between dementia \cite{Levakov2020, MONTAVON20181}.

In recent years, a new field of research dedicated to the explanation of deep learning models has emerged: Explainable Artificial Intelligence (XAI) \cite{xai1, xai2,xai3}. In that field, heatmaps have emerged as a popular visualization tool to interpret Deep Learning models working on images. A heatmap indicates what part of an input image contributes the most to a deep network output \cite{saliency, CAM, zhang2018}. In other words, a heatmap reflects the importance of imaging features extracted from an image by a deep neural network to support its decision, and how much local image patterns contribute to these important features. The first heatmap method proposed, coined saliency maps, was produced by back-propagating the gradients of a network output through all the layers of the model until reaching the input layers \cite{saliency}. This core idea was improved and generalized several times in the following years, such as in the Guided Backpropagation method that builds on the Deconvolution network \cite{deconv} and where only positive gradients are propagated through ReLU network layers \cite{GB,relu}. In the Class Activation Maps (CAM), network activations are considered instead of back-propagated gradients \cite{CAM}. The Integrated Gradient (IG) approach consists of averaging gradient maps generated from multiple scaled inputs \cite{IG}. In the Layer-wise Relevance Propagation (LRP) method, a set of preservation rules are applied when back-propagating a network's activations and, in particular, treating positive and negative neural network activations in different ways \cite{LRP, LRP2, LRPbetaRule}. These strategies are also implemented in the DeepLift method \cite{DeepLift}, where baseline activations are subtracted from neuron activations during the propagation. These baseline activations are generated by passing task-specific reference images to the networks \cite{DeepLift}. The most recent approaches combine multiple methods to generate fine visualizations that can be produced at different network depths, such as the Guided Grad CAM method that combines guided back-propagated gradients with class-activation maps generated from output gradients \cite{GB, CAM, GGC}. Heatmaps methods can be selected based on their implementation invariance \cite{IG}, their robustness for input perturbations \cite{input_perturb}, model weight randomization \cite{sanity_checks}, and the relevant information they capture in the saliency maps they produce \cite{saliency_entropy}. In the studies where no ground truth is available to estimate the quality of the heatmaps, this evaluation is particularly difficult to conduct \cite{LRP_AD}.

\begin{table*}[t]
\centering
\begin{tabular}{lgcg}
& controls & AD & p-value \\
\hline
Study participants (n) & 252 & 250 & \\
%Cohort (ADNI 1/2/3) & 92/160/0 & 78/138/34 & \textless 0.001 \\
MRI B0 field (1.5T/3T) & 92/160 & 78/172 & 0.22 \\
Sex (M/F) & 131/121 & 143/107 & 0.25 \\ 
Mean age (std) & 74.41 (6.00) & 74.93 (8.01) & 0.41 \\
Mean MMSE (std) & 29.06 (1.25) & 22.95 (2.23) & \textless 0.001 \\
Mean years of education (std) & 15.25 (2.95) & 16.35 (2.65) & \textless 0.001 \\
\hline
\end{tabular}
\caption{\label{table1} ADNI participants selected in this work. Fisher's exact test detected no significant difference in the proportion of men in the two groups (p=0.25) and no significant difference in the proportion of scans acquired with a 3 Tesla MRI scanner (p=0.22). The T-test detected no significant mean age difference (p=0.41), while mini-mental state examination (MMSE) values \cite{MMSE} were significantly worse in the AD group. Education information was only available for 220 AD study participants and 211 controls and indicated significantly longer education in the AD group.}
\end{table*}

In the neuroimaging field, clinical studies have been conducted for decades to establish how brain dementia affects the brain \cite{vbm, ashburner2001vbm}. A considerable number of voxel-based morphometry studies (VBM) have been conducted to discover which brain atrophies observed in the aging brain can be imputed to an underlying Alzheimer's disease \cite{ad_vbm1, ad_vbm2, ad_vbm3, ad_vbm4, ad_vbm5}. When a meta-analysis is conducted, these VBM studies are often summarized into a single brain map indicating what brain regions are affected by the disease \cite{meta1,meta2,meta3}. VBM studies capture the univariate significance of local tissue changes: they indicate how much a brain disorder such as AD has impacted local brain tissues. This `decoding' approach reverses the `encoding' approach adopted by neural networks, where local tissue changes are aggregated into non-linear features used to predict patient diagnosis. However, under the assumptions that AD only affects localized brain regions \cite{ad_vbm1, ad_vbm2, ad_vbm3, ad_vbm4, ad_vbm5} and that neural networks focus on a restricted set of relevant brain regions when diagnosing AD, VBM and heatmaps should overlap to highlight brain regions associated with and predictive of the disease. Since the different heatmap methods capture imaging pattern contributions in various ways, the overlap between heatmaps and Alzheimer's disease VBM patterns is also expected to depend on the heatmap calculation; some methods focusing on high-level features are more difficult to relate to voxel-wise VBM results. Lastly, it is unclear how well heatmaps derived from a restricted data set would replicate in larger AD neuroimaging cohorts and if that overlap between univariate significance and neural network features' importance would be preserved. 

As far as we know, none of these questions have been explored so far. We propose to address them at the same time, in this work, by quantifying the amount of overlap that can be reached between a "ground truth" univariate significance map provided by a large VBM meta-analysis and heatmaps derived by the most advanced methods from a convolutional neural network achieving state-of-the-art classification performance for Alzheimer's disease classification on an independent sample of MRI scans. More specifically, we evaluate the ability of three prominent CNN heatmap methods, the Layer-wise Relevance Propagation (LRP) method \cite{LRP}, the Integrated Gradients (IG) method \cite{IG}, and the Guided grad-CAM (GGC) \cite{GGC} method, to capture Alzheimer's disease effects by training 3D CNN classifiers using T1-weighted MRI scans part of the ADNI data set, and measuring the overlap between their heatmaps and a binary brain map derived from a meta-analysis of voxel-based morphometry studies conducted on other T1 MRI scans. Figure\ref{overview} summarizes our approach.

\section{Materials and Methods}
\subsection{ADNI Study Participants}
A total of 502 ADNI participants were included in this study, 250 participants were diagnosed with AD and 252 controls. 170 participants were part of the ADNI1 study (92 controls, 78 AD), 298 were enrolled in the ADNI2 study (160 controls, 138 AD), and the last 34 participants were recruited for ADNI3 (0 controls, 34 AD). Study participant demographics are reported in Table \ref{table1}.

\subsection{ADNI Data and Processing}
For each participant, a raw structural T1-weighted MRI scan was downloaded from the Alzheimer’s Disease Neuroimaging Initiative (ADNI) database (adni.loni.usc.edu). As a preparation for the present study, the scans were further processed as follows. 

First, the multi-atlas brain segmentation pipeline (MUSE) was used for skull-stripping the T1-weighted MRI scans and generating a gray matter map \cite{muse}. This automated processing pipeline starts by denoising the T1 scans using the N4 bias field correction \cite{n4} provided as part of the Advanced Normalization Tools software library (ANTs, version 2.2.0) \cite{ants}. Then, the denoised scans are registered using ANTS nonrigid SyN registration \cite{ants,antsSyn} and DRAMMS \cite{dramms} to a set of 50 brain atlases where brain masks have been manually segmented. These registrations are used to warp the atlas brain masks into the space of the T1 scan to process, where they are combined by majority voting to produce an accurate brain mask \cite{muse}. The brain is then segmented into white matter, gray matter, and cerebrospinal fluid using FSL FAST (version 5.0.11) \cite{fsl}, and parcellated into regions of interest by registering a set of 50 manually segmented brain atlases \cite{muse}. 

Then, the skull-stripped T1 scans produced by MUSE were registered to the 1 mm resolution 2009c version of the ICBM152 MNI atlas \cite{mni1,mni2,mni3} using the non-rigid registration method SyN part of the ANTs library (ANTs version 2.3.4) \cite{ants,antsSyn}. The deformation field produced by ANTs was applied, after masking out white matter and ventricles using the tissue maps generated by MUSE, to bring only the gray matter into the MNI space. The same transformation was applied to move the MUSE grey matter mask to the MNI space. Lastly, each T1 scan was normalized individually, by dividing the T1 intensities by the maximum intensity within the brain. 

\subsection{Convolutional Neural Networks}

\begin{figure*}
\includegraphics[width=1.0\linewidth]{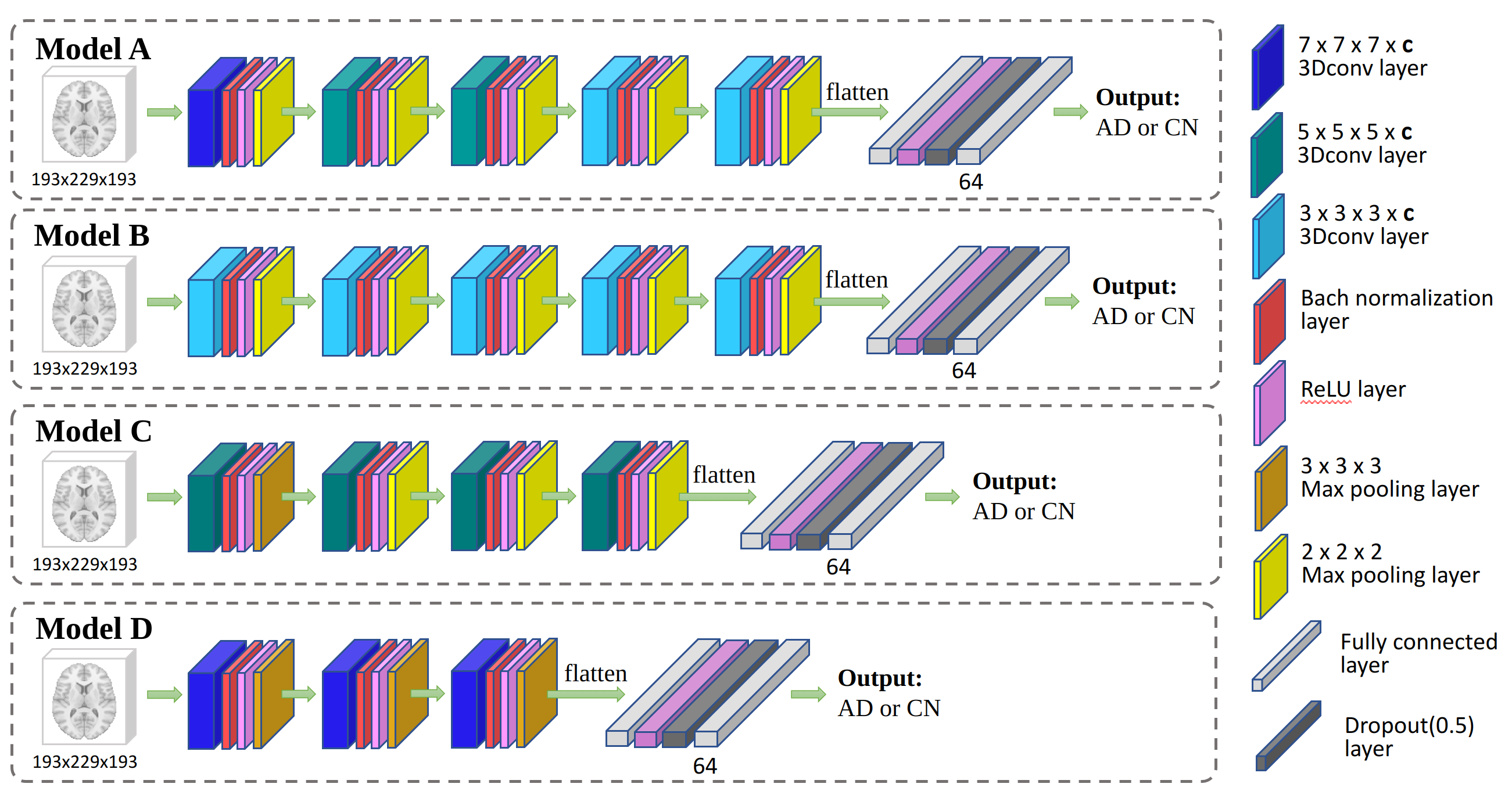}
\caption{\label{architecture}CNN models were used in this study to classify ADNI AD participants and controls. For each architecture, several numbers of channels \textbf{c} were tested.}
\end{figure*}

A convolutional neural network (CNN) consists of a set of convolutional layers applying convolution operators to gradually condense input data into a set of high-level features that are passed through fully connected layers to produce the final output of the network \cite{alexnet}. This output is usually a single value scaled between 0 and 1 when the CNN is used for binary classification \cite{alexnet}. Convolutional layers are often combined with the ReLU activation layer filtering negative outputs \cite{relu}, max-pooling layers reducing the dimension of the data \cite{alexnet}, and batch normalization layers helping the neural network model optimization \cite{BN}. 

In this work, four different 3D CNN architectures of varying complexity were compared. The first architecture, which will be referred to as Model A, was made of five convolutional layers with decreasing kernel sizes: one layer with a kernel size of $7\times 7\times 7\times c$, two layers with kernel sizes of $5\times 5\times 5\times c$, and two layers with kernel sizes of $3\times 3\times 3\times c$, where $c$ denotes the number of channels and it was fixed for each CNN separately. Model B had five convolutional layers with the same kernel size of $3\times 3\times 3\times c$. Model C had four convolutional layers with the same kernel size of $5\times 5\times 5\times c$. Model D was made of three convolutional layers with the same kernel size of $7\times 7\times 7\times c$. All convolutional layers were followed by a batch normalization layer, a ReLU activation layer \cite{relu}, and a max-pooling layer \cite{alexnet}. For each architecture, the number of channels varied from 24 to 52 to build networks of increasing numbers of parameters. On top of these convolutional layers, all the CNNs were completed by two fully connected layers separated by a ReLU layer \cite{relu} and a dropout layer fixed to 0.5 to prevent overfitting \cite{dropout}. The first fully connected layer was obtained by flattening the features produced by the last convolutional layer. The second layer was set to contain 64 neurons and to produce a continuous output corresponding to the AD diagnosis. These CNN architectures are summarized in Figure \ref{architecture}. 

These networks were trained to distinguish between the MRI scans of AD and control in our ADNI data set. Cross-entropy was used as a loss function for the classification, and that loss was minimized using the Adam optimizer with a learning rate of 0.0001 and a weight decay of 0.0001 \cite{adam}. CNN accuracy was evaluated via 5-fold cross-validation, by splitting the data set five times into a training set of 322 scans, a validation set of 80 scans, a test set of 100 scans for the first four folds and a training set of 322 scans, a validation set of 78 scans, a test set of 102 scans for the last fold. An early stopping criterion was implemented by monitoring the validation loss and forcing the training process to stop after 10 epochs producing no improvements in the validation loss. These CNN architectures and their optimizers were selected to match standard architectures and their default optimization parameter values and, in particular, state-of-the-art Alexnet \cite{alexnet} and Google net \cite{googlenet}.

All the models were trained on a high-performance computing system equipped with Nvidia v100 GPUs. The training required 32GB of RAM and was completed within 3 hours to 8 hours for each fold, depending on the model complexity. The proposed network was built with Pytorch \cite{pytorch}. 

The classification accuracy obtained for all the CNNs tested was compared with the classification accuracy of linear SVMs with the following set of C parameters \cite{svm, scikit-learn}: $10^{-6}$, $10^{-5}$, $10^{-4}$, $10^{-3}$, $10^{-2}$, $10^{-1}$, $1$, $10$, $10^{2}$, $10^{3}$, $10^{4}$. The CNN model with the best cross-validated accuracy was then retained to compute heatmaps highlighting the brain regions selected by the model to distinguish AD and control brains. 

\subsection{CNN Heatmap Methods}
In this study, three prominent CNN heatmap methods are selected, the Layer-wise Relevance Propagation (LRP) method \cite{LRP}, the Integrated Gradients (IG) method \cite{IG}, and the Guided Grad-CAM (GGC) \cite{GGC} method. These methods were used to produce a heatmap for each test scan and then averaged to produce a single heatmap for each heatmap method indicating what brain regions were used the most by the selected CNN when classifying the brain scans to distinguish AD and control ADNI participants \cite{LRP, IG, GGC}.

The Layer-wise Relevance Propagation (LRP) method produces a heatmap by estimating a relevance score for each pixel of input passed to a CNN model. The relevance score is computed by propagating CNN outputs backward in the network according to a specific set of rules \cite{LRP}. These rules are designed to preserve relevance scores from layer to layer and are often modified to reduce the noise in relevance scores, improve their sparsity, or treat positive and negative neural network activations differently \cite{LRP, LRPbetaRule, LRP2}. In this work, we used the $\beta$-rule LRP algorithm implemented by B{\"o}hle et al. \cite{LRPbetaRule, LRP_AD}. The $\beta$ parameter aiming at balancing the relevance scores associated with positive and negative neural network activations was set to 0.5 to account for both activations in a similar manner \cite{LRP}.

The Integrated Gradients (IG) method was introduced to guarantee two desirable heatmap properties: sensitivity and implementation invariance \cite{IG}. Sensitivity refers to the ability of a heatmap method to produce null relevance scores for network inputs that are not contributing to the network output. Most of the methods published before IG either did not satisfy the sensitivity requirement, such as Guided Backpropagation\cite{GB}, Deconvolution networks \cite{deconv}, and DeepLift \cite{DeepLift}, or were not invariant to the neural network implementation, such as LRP \cite{LRP} and DeepLift \cite{DeepLift}. The Integrated Gradients (IG) method produces the heatmap of an input image by multiplying the input with a scaling factor uniformly selected between 0 and 1 several times in a row, and computing the gradient for each scaled input via backpropagation, and then averaging these gradients \cite{IG}. The IG implementation used in this work will be available on \href{https://github.com/UTHSCSA-NAL/CNN-heatmap}{https://github.com/UTHSCSA-NAL/CNN-heatmap/}.

Guided Grad-Cam (GGC) combines Grad-CAM and Guided Backpropagation \cite{GB, GGC}. Grad-CAM is a generalization of the Class Activation Mapping (CAM) \cite{CAM} that can be implemented for any CNN without model changes or re-training. Grad-CAM generates a heatmap for a CNN layer by applying a ReLU function \cite{relu} to a linear combination between the activations obtained at that layer and the backpropagated gradients from subsequent layers \cite{GGC}. In Guided Grad-CAM, these Grad-CAM heatmaps are up-sampled to the resolution of the input data and element-wise multiplied with a heatmap generated by Guided Backpropagation to produce heatmaps with the same resolution as the input data \cite{GB, GGC}. During our experiments, we only considered Guided Grad-CAM heatmaps based on the Grad-CAM maps computed for the last convolutional layer of our CNNs, as suggested in the original GGC publication \cite{GGC}. GGC was implemented as part of the Captum PyTorch library \href{https://captum.ai/}{https://captum.ai/}.

\subsection{Meta-analysis ALE Map} % AD+MCI meta-analysis
\begin{table}
\centering
\begin{tabular}{|c|g|c|g|}
\hline
 & n & men/women & mean age \\
\hline
CN & 2118 & 853/1127 & 69.9 \\
AD/MCI & 1699 & 725/882 & 71.6 \\
\hline
p-value & & 0.224 & 0.03 \\
\hline
\end{tabular}
\caption{\label{ALEmapsTable} The meta-analysis ALE map summarizes 77 VBM studies published in 58 articles. Gender information was missing in 3 MRI VBM studies. A Fisher exact test was conducted to compare men/women counts, and an unpaired T-test was conducted to compare group mean ages. After Bonferroni correction for two tests, none of the differences observed was significant at level p=0.05.}
\end{table}

The meta-analysis map was produced by reprocessing a set of voxel-based morphometry (VBM) studies collected from a prior meta-analysis \cite{vbm}. We produced a brain map by applying the activation likelihood estimation (ALE) method \cite{ale, ale2, ale3, ale4} implemented in the GingerALE software (version 3.0.2) \cite{gingerAle, gingerAle2} to combine the selected VBM studies into a single ALE map indicating what atrophies observed in the brain were likely to be associated with Alzheimer's disease \cite{ale}. More specifically and following \cite{ale5, muller2}, GingerALE was running for a cluster-forming p-value of 0.001 and a cluster-level significance level of 0.05. Cluster significance was estimated by conducting a thousand random permutations. The continuous map generated by GingerALE, where all non-significant brain locations had been assigned null values, was thresholded at its smallest non-zero value to produce a binary map suitable for a comparison with the thresholded CNN heatmaps. 

\subsection{Evaluation Metrics}

The ability of the CNN heatmap methods presented in the previous section to capture brain alterations associated with Alzheimer's disease was estimated by measuring the Dice overlap between binary maps obtained by smoothing and thresholding the heatmaps with the binary brain map derived from a large meta-analysis that summarized the brain regions affected by Alzheimer's disease in T1 MRI scans.

More specifically, CNN heatmap values were replaced by their absolute values. The heatmaps were then smoothed by sixteen different Gaussian kernels of full width at half maximum (FWHM) ranging from 1 mm to 32 mm (1 mm, 2 mm, 3 mm, 4 mm, 5 mm, 6 mm, 7 mm, 8 mm, 9 mm, 10 mm, 12 mm, 16 mm, 20 mm, 24 mm, 28 mm, 32 mm). For each smoothed heatmap and the original heatmaps, 50 values were evenly selected between the minimum and the maximum heatmap value to threshold the heatmaps. The reason to smooth a heatmap is to make it comparable to a meta-analysis map since the ALE algorithm inherently adds Gaussian smoothing to the locations of reported foci. The 850 binary maps obtained in this way were compared with the meta-analysis map by computing a Dice overlap. This approach was chosen to explore and mitigate spatial resolution discrepancies between the CNN heatmaps and the meta-analysis map.

In addition to the Dice overlaps, Receiver Operating Characteristic (ROC) curves and precision-recall curves were computed to compare each smoothed heatmap with the meta-analysis map, by considering the meta-analysis map as a set of reference binary labels. The area under the curve was reported as well. These additional measures of overlap between heatmaps and the meta-analysis map were used to validate the findings derived from Dice overlaps. 

\subsection{Additional Synthetic Validation}
The methods presented in this work were validated by processing two synthetic data sets. The first data set, the ``single-subject" data set, made of 10000 images, was generated from a single healthy control subject MRI scan (mean MRI intensity = 2600, std = 756) from our ADNI data set and was downsampled to a size of $65\times 77\times 65$ voxels to reduce the computational burden. In half of the images, the MRI intensity in the hippocampus regions was increased by a random value ranging between 0 and 2500 and simulating a disease effect on grey matter tissue. Then, Gaussian noise (mean=0, std=2000) was added to all synthetic images, and a Gaussian smoothing of 4 mm FWHM was applied. The second data set, the ``whole-cohort" data set, also made of 10000 images, was generated using 250 healthy controls from our ADNI data set and downsampled to $65\times 77\times 65$ voxels. Each healthy control scan was used to generate 40 images. In half of these images, the MRI intensity in the hippocampus regions was increased by a random disease effect between 0 and 2500. Then, Gaussian noise (mean=0, std=2000) was added to all synthetic images, and a Gaussian smoothing of 4 mm FWHM was applied. 

Eleven linear SVMs were trained to distinguish the synthetic scans with and without disease effect, for the parameters tested with the clinical data (C parameter in $10^{-6}$, $10^{-5}$, $10^{-4}$, $10^{-3}$, $10^{-2}$, $10^{-1}$, $1$, $10$, $10^{2}$, $10^{3}$, $10^{4}$) \cite{svm, scikit-learn}. Then, four CNN models similar to the models shown in Figure \ref{architecture} were trained: a Model A containing one $7\times 7\times 7\times c$ and one $5\times 5\times 5\times c$ convolutional layers, where the number of channels $c$ was set to $4$ (ModelA4l2), a Model B containing two $3\times 3\times 3\times c$ convolutional layers (ModelB4l2), a Model C containing two $5\times 5\times 5\times c$ convolutional layers (ModelC4l2), and a Model D containing two $7\times 7\times 7\times c$ convolutional layers (ModelD4l2). The number of convolutional layers in these CNNs was reduced, compared to the original CNN architectures used for classifying ADNI scans, to fit the size of the downsampled synthetic images. They were trained for a cross-entropy classification loss, that was minimized using the Adam optimizer with a learning rate of 0.0001 and a weight decay of 0.0001 \cite{adam} for a hundred epochs. Please refer to Figure \ref{architecture} for their detailed architecture. For each data set, LRP, IG, and GGC were used to compute a heatmap for the CNN model reaching the best five-fold cross-validated accuracy. The coefficients of the best SVM and the heatmap values were then compared with the binary map of the hippocampus by computing Dice overlaps at different spatial smoothing levels as explained in the previous Sections.

\section{Results}
\begin{figure*}[!t]
\centering
A\includegraphics[width=0.48\linewidth]{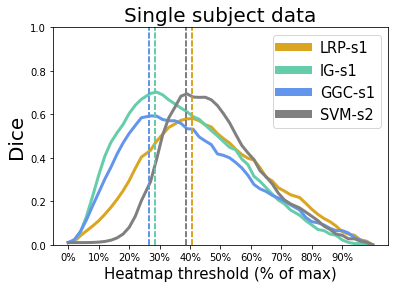}
B\includegraphics[width=0.48\linewidth]{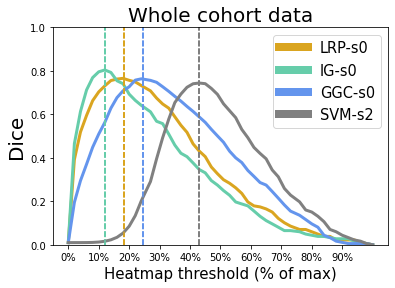}
\caption{\label{dice_synthetic} Dice curves reaching the best overlap between the brain region altered in the synthetic data and heatmaps or SVM coefficients attempting to capture that region; (A) for the single-subject data set, and (B) for the whole-cohort data set. s0 indicates a heatmap that was unsmoothed, s1 indicates a heatmap smoothed by a Gaussian kernel of 1 mm FWHM, and s2 indicates a 2 mm FWHM smoothing.}
\end{figure*}

\subsection{Classification performance in synthetic data}
For both synthetic data sets, the model D4l2 reached the best five-fold cross-validated accuracy, with 91\% accuracy for the single-subject data and 90.1\% for the whole-cohort data, and systematically outperformed the best SVM models, that were obtained for both data sets by setting $C=0.1$ (87\% accuracy for single-subject data set and 87.5\% accuracy for whole-cohort data set). More specifically, for the single-subject data set, model A4l2 also outperformed the best SVM, with an accuracy of 90\%, but Model C4l2 and B4l2 produced worse classification results, with respectively 81\% and 74\% accuracy. For the whole-cohort data set, on the contrary, B4l2 was the second-best model with 90.06\% accuracy, followed by A4l2 (89.9\%) and C4l2 (87.9\%) and all CNN models were more accurate than the best SVM tested. 

\subsection{Heatmaps derived from synthetic data}

The Dice overlaps between heatmaps and the binary hippocampus map are shown in Figure \ref{dice_synthetic}. All smoothing results are reported in Supplementary materials (Section 1). For the single-subject data set, the best Dice overlap measured for LRP, IG, GGC, and SVM is 0.581 when the LRP heatmap was smoothed by a 1 mm FWHM Gaussian smoothing, 0.703 for the IG heatmap with a 1 mm smoothing, 0.593 for GGC heatmap with a 1 mm smoothing and 0.694 for the SVM heatmap with a 2 mm smoothing. For the whole-cohort data set, the best Dice overlap measured for LRP, IG, GGC, and SVM is 0.766 when the LRP heatmap without Gaussian smoothing, 0.804 for the IG heatmap without smoothing, 0.763 for GGC without smoothing, and 0.744 for the SVM heatmap with a 2 mm smoothing. Their corresponding heatmaps achieved the best overlap with the hippocampus map and the heatmaps thresholded at 5\% of their maximum values are shown in Supplementary materials (Section 1). Those plots indicate that IG heatmaps have a better focus on the hippocampus than GGC and LRP heatmaps. 

These results demonstrate the ability of CNN heatmaps to capture localized and specific brain alterations on a synthetic data set and, in particular, when the hippocampus is affected similarly as in real clinical data. IG heatmaps reached the best overlaps for both data sets and required the smallest spatial smoothing to achieve that overlap (1 mm for the single-subject data set, no smoothing for the whole-cohort data set).

\subsection{Classification performance in ADNI}
The 5-fold cross-validation accuracy of all the CNN and SVM models tested during this work is reported in table \ref{cnn_performances}. The best CNN accuracy was achieved by the Model B with 44 channels (ModelB44) and reached 87.24\%. This accuracy is six percent better than the cross-validated accuracy obtained with the best SVM model(c= 0.001), which is close to 81.2\% and that was obtained through grid search for a set of c parameters : $10^{-6}$, $10^{-5}$, $10^{-4}$, $10^{-3}$, $10^{-2}$, $10^{-1}$, $1$, $10$, $10^{2}$, $10^{3}$, $10^{4}$.

\begin{table}[!htb]
\begin{center}
\begin{tabular}{|c|c|c|c|c|}
\hline
c & Model A & Model B & Model C & Model D \\
\hline
$24 $ & $76.41\%$ & $82.65\%$& $75.47\%$& $58.59\%$ \\
\hline
$28 $ & $83.85\%$ & $84.45\%$ & $65.59\%$ & $54.78\%$\\
\hline
$32 $ & $78.86\%$ & $85.44\%$ & $84.45\%$ & $58.20\%$ \\
\hline
$36 $ & $83.84\%$ & $86.05\%$ & $72.18\%$ & $55.80\%$ \\
\hline
$40 $ & $78.05\%$ & $86.25\%$ & $58.18\%$ & $47.21\%$ \\
\hline
$44 $ & $61.80\%$ & $\mathbf{87.25\%}$ & $55.65\%$ & $52.99\%$ \\
\hline
$48 $ & $59.81\%$ & $86.45\%$ & $64.45\%$ & $55.48\%$ \\
\hline
$52 $ & $61.41\%$ & $86.46\%$ & $62.64\%$ & $47.81\%$\\
\hline
\hline
\multicolumn{5}{|c|}{best SVM: C=0.001, accuracy: $81.19\%$} \\
\hline
\end{tabular}
\end{center}
\caption{\label{cnn_performances} 5-fold cross-validation accuracy for the classification of ADNI participants, for all CNN architectures, all numbers of channels $c$, and the best Linear SVM. The number of channels corresponds to the number of 3D convolutional kernels used in each CNN convolutional layer.}
\end{table}

\subsection{Meta-Analysis ALE Maps}
\begin{figure*}
\includegraphics[width=1.0\linewidth]{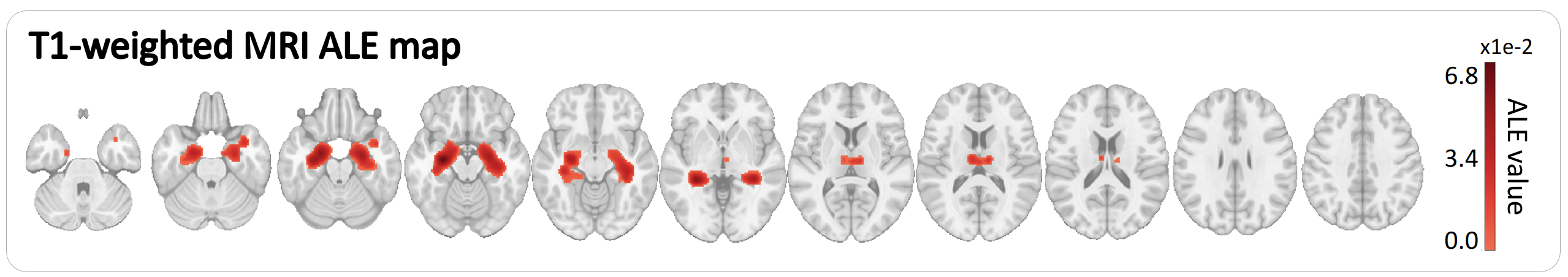}\
\centering
\caption{\label{ALEmaps} Meta-analysis ALE maps in the MNI152 template space. }
\end{figure*}

Figure \ref{ALEmaps} presents the ALE map used in this work. The structural MRI ALE map summarizes 77 neuroimaging studies reporting 773 locations in the brain affected by Alzheimer's disease, discovered by analyzing the neuroimaging data of a total of 3817 study participants around the world (2118 controls, 1699 MCI or AD). The complete list of publications combined in this map is reported in Supplementary materials (Section 2). 

\subsection{Overlaps between ADNI CNN heatmaps and the ground truth derived from the meta-analysis}
The five ModelB44 models trained during the 5-fold cross-validation were used to generate heatmaps. For each model, a heatmap was generated for each test scan with each of the three heatmap methods: LRP, IG, and GGC. The 502 individual heatmaps obtained for each method were averaged into a single heatmap that was compared with the binary meta-analysis map to evaluate the performance of the method. In addition, an SVM coefficient map was derived by training a linear C-SVM using all the data and for the C parameter producing the best cross-validated accuracy and retaining the weight of this model to create a brain map. 

\begin{figure}
\centering
\includegraphics[width=1.0\linewidth]{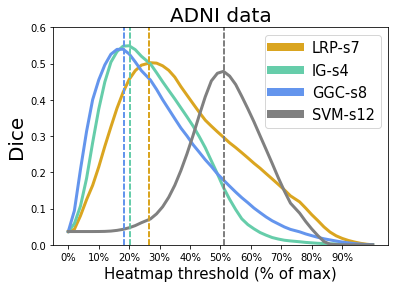}
\caption{\label{dice} For each heatmap method, the Dice curve of ADNI data corresponding to the spatial smoothing reaches the best overlap with the meta-analysis. LRP-s7: LRP heatmap smoothed by a Gaussian kernel of 7 mm FWHM, IG-s4: IG heatmap smoothed by a 4 mm FWHM smoothing, GGC-s8: GGC heatmap smoothed by an 8 mm smoothing, SVM-s12: SVM coefficient map smoothed by a 12 mm FWHM smoothing.}
\end{figure}

\begin{figure*}
\includegraphics[width=1.0\linewidth]{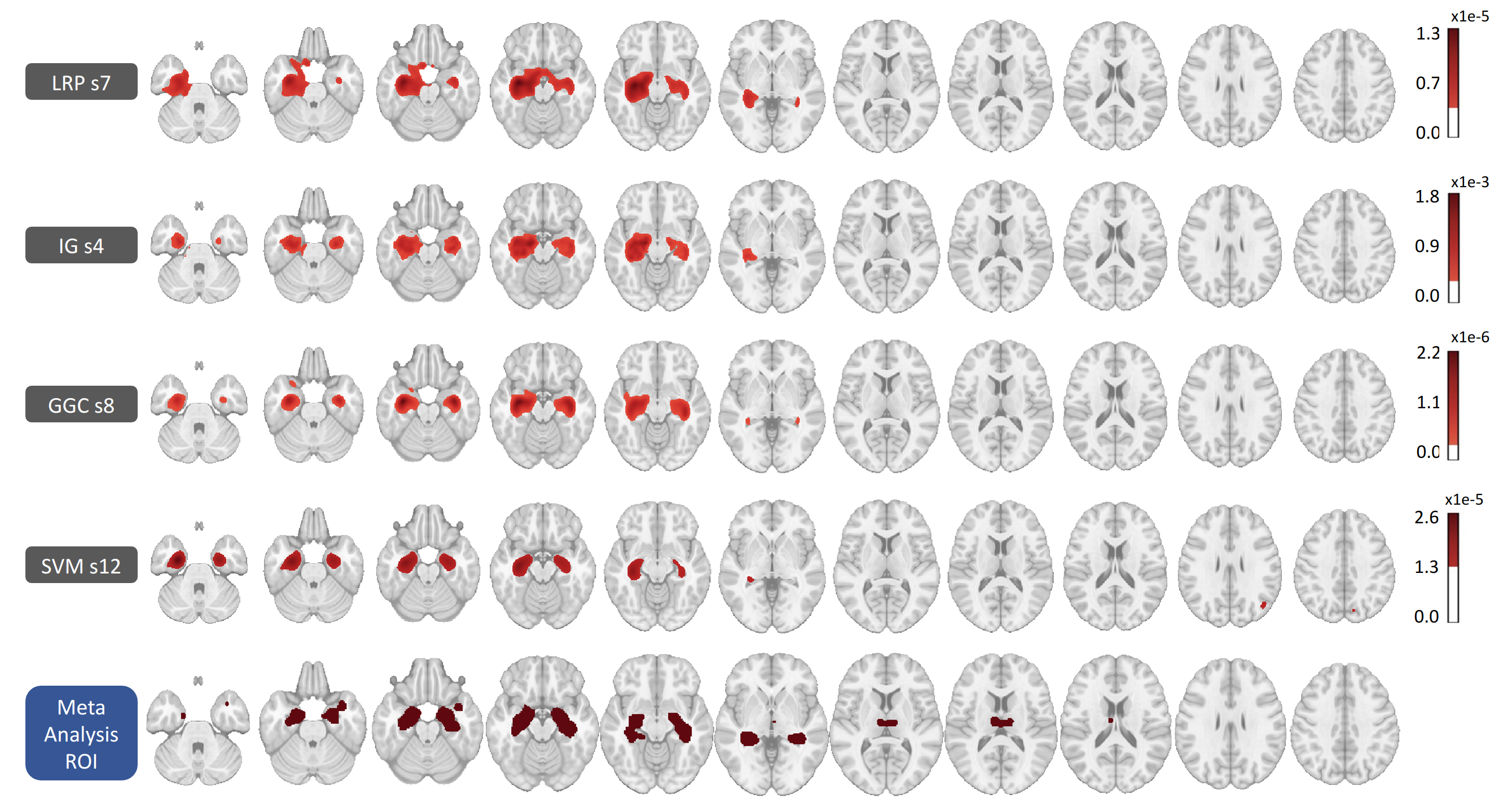}
\caption{\label{plot}Heatmaps corresponding to the best Dice overlap with the meta-analysis map, for all the CNN heatmaps methods tested in this work and the best linear C-SVM. The meta-analysis map is binary.}
\end{figure*}

\begin{figure*}
\includegraphics[width=1.0\linewidth]{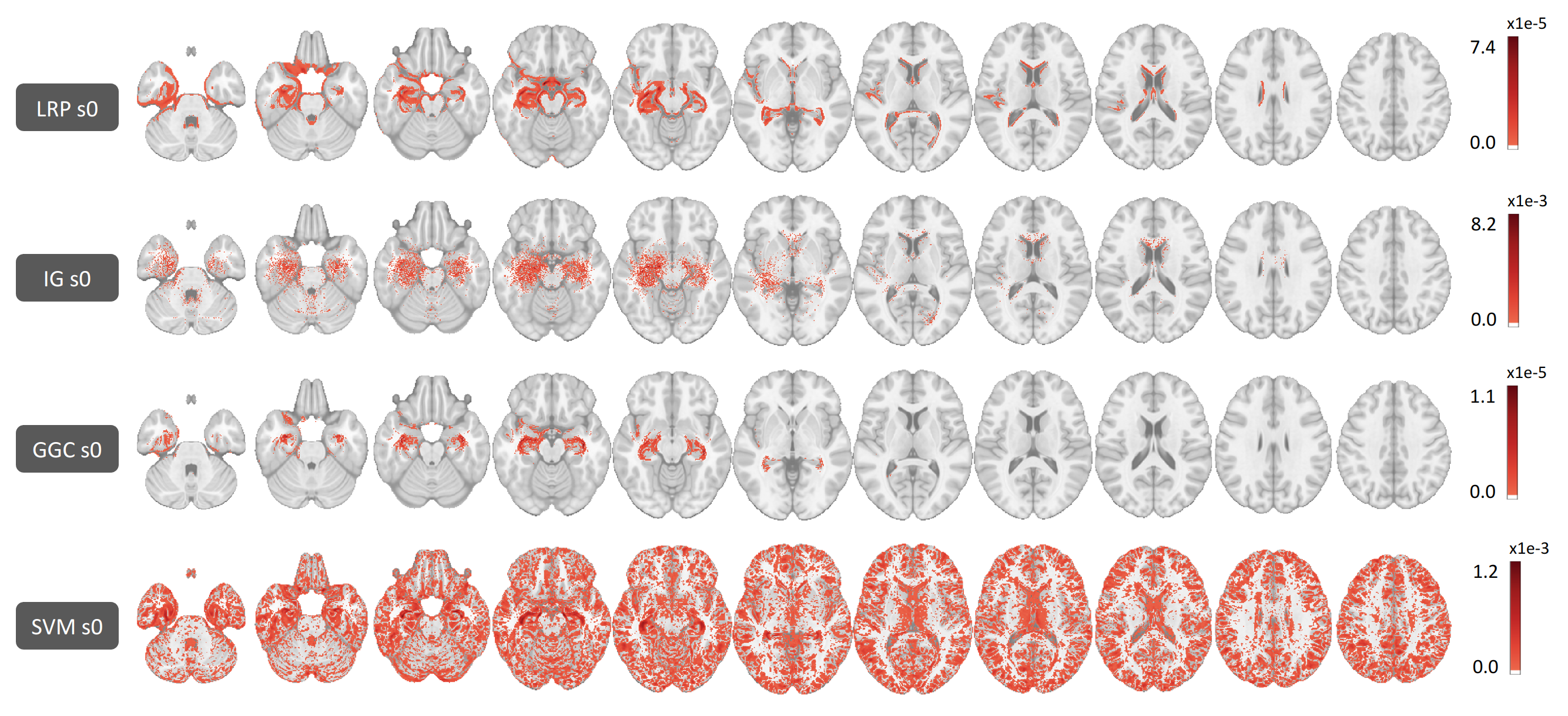}
\caption{\label{plot2}Heatmaps without smoothing, thresholded at 5\% of their maximum value.}
\end{figure*}

Figure \ref{dice} reports all the Dice measured between the heatmaps and the meta-analysis map. The best Dice overlap measured for LRP, IG, GGC, and SVM was 0.502 when the LRP heatmap was smoothed by a 7 mm FWHM Gaussian smoothing, 0.550 for the IG heatmap with a 4 mm smoothing, 0.540 for GGC with an 8 mm smoothing and 0.479 for the SVM heatmap with a 12 mm smoothing. The heatmaps with the best overlaps with the meta-analysis are shown in Figure \ref{plot}. Figure \ref{plot2} displays the unsmoothed heatmaps. The LRP heatmaps select more regions than the other maps and appear to be noisier. On the other hand, IG heatmaps have a better focus on the regions highlighted by the meta-analysis, but the unsmoothed IG map presents an unrealistic scatter. IG produced a map that was simultaneously more relevant than the LRP heatmap and less scattered than the GGC heatmap. In comparison, the unsmoothed SVM heatmap covers most of the grey matter. The linear SVM produced slightly larger weight amplitudes in the regions relevant for the diagnosis, but an aggressive smoothing was required to make this effect emerge in Figure \ref{plot}. 

\section{Discussion}
In the present study, we reported the first data-driven validation, for the study of Alzheimer's disease, of three prominent CNN heatmap methods: Layer-wise Relevance Propagation (LRP), Integrated Gradients (IG), and Guided Grad-CAM (GGC). The heatmaps produced by these methods, for a CNN classifier producing the best AD classification among a large set of CNN architectures tested using ADNI T1-weighted MRI scans, were compared with a binary meta-analysis ALE map obtained by combining 77 Alzheimer's disease VBM studies. Our results indicate that the CNN heatmaps captured brain regions that were also associated with AD effects on the brain in the meta-analysis. 

\subsection{Best deep learning-based classification model}
The best 5-fold cross-validation accuracy (87.24\%) was obtained for a Model B with 44 channels. Overall, Model B accuracy was stable when the number of channels was varied, varying only between 83\% and 87\%. Model A and C were less stable: their accuracy ranged between 60\% and 84\% and 63\% and 84\% respectively. Model D produced only poor classifications, for an accuracy ranging between 47\% and 59\%. We think that these differences can be explained by overfitting, as we noticed that Models D usually contain more trainable parameters than Models C of a similar number of channels. Models C usually contain more parameters than Models A, and Models B are the smallest models. For 44 channels, for instance, Model D required the training of 1569058 parameters, Model C 957502 parameters, Model A 705866, and Model B only 436410 parameters. 

\subsection{Explainable AI and neuroimaging}

The neuroimaging field has developed meta-analysis brain maps to summarize domain knowledge \cite{fox2005brainmap, vanasse2018brainmap}, which we use to evaluate the CNN heatmaps. Contrary to prior studies, where the quality of heatmaps was visually or qualitatively assessed \cite{input_perturb, LRPbetaRule, heatmap_review, heatmap_tau}, we obtained precise quantitative measures by computing overlaps with a ground-truth map derived from a large-scale meta-analysis. We explored a broad range of heatmaps' spatial smoothing intensities and we found that the heatmaps overlapped the most with the meta-analysis for Gaussian smoothing kernels between 4 mm and 8 mm FWHM. These Gaussian kernels are similar to the kernels usually applied by GingerALE when producing meta-analysis maps \cite{gingerAle, gingerAle2}.

\subsection{Evaluation experiments for the heatmaps derived from deep learning models}

For all the CNN heatmap methods, the best heatmaps indicated that changes in the hippocampus regions in both hemispheres were a crucial pattern during the classification of ADNI participants with AD and healthy controls. These results are perfectly in line with the literature, where the effect of Alzheimer's disease on the hippocampus has been well-characterized \cite{habes2016BA, hippo1, hippo2, hippo3, hippo4}. We obtained moderately good Dice overlaps between heatmaps and the meta-analysis ground-truth, ranging from 0.5 for the best heatmap generated by the LRP method to 0.55 for the best IG heatmaps. 

Direct analysis of the heatmaps, without spatial smoothing, established that all CNN heatmaps were better at focusing on relevant brain regions than linear SVM weights. IG and LRP produced scattered heatmaps that benefited the most from spatial smoothing, gaining up to 0.19 and 0.18 in Dice overlap with the ground truth as the size of the Gaussian kernels was varied. GGC Dice overlap was only improved by 0.16 at most. In comparison, the SVM weight map was so scattered and noisy that a Dice improvement larger than 0.3 was observed when the map was smoothed. We refer the readers to the Supplementary materials for the complete set of Dice overlaps measured during this experiment (Section 3). The LRP heatmap was the noisiest, and produced the least symmetric results by selecting more voxels in the left hemisphere, as reported in prior studies \cite{LRP_AD}. 

IG produced the heatmaps with the largest overlaps with the meta-analysis, and that overlap required less spatial smoothing. These results suggest that the IG heatmap while being more scattered than other heatmaps, was overall less noisy.  All heatmap methods produced brain maps closer to the meta-analysis map than the map derived from the baseline support vector machine and were better focused on brain regions impacted by the disease than the SVM map. The additional overlap measures presented in Supplementary materials (Section 4) also indicate a better overlap between the IG heatmaps and the meta-analysis ground truth, and these results are in line with the synthetic results, where IG also outperformed other heatmap methods.

\subsection{Data augmentation}

Various techniques could be used to improve classification performance, such as data augmentation, which is used to enhance performance by enlarging the training set \cite{DEEPMIRAD}. In this work, we did not employ data augmentation as we were aiming to capture biology-informed patterns, and the most standard form of data augmentation, the inclusion of translated and rotated copies of the training scans \cite{DEEPMIRAD}, would have blurred the boundaries of the brain regions that the CNN heatmaps were aiming to capture. 
In the future, we will check whether the use of more advanced data augmentation methods, such as the introduction of realistic noises that preserve the boundaries between grey matter and white matter in the MRI scans, could be used to carry out a data augmentation that retains the boundaries of the regions of interest.

\subsection{Evaluation Metrics}

Multiple metrics could be used to measure the overlap between the binary meta-analysis map provided by GingerALE, the continuous SVM coefficients maps, and the continuous brain maps generated by the heatmap methods. In this work, we decided to threshold the absolute value of the continuous heatmaps and we used a well-established metric to measure the overlap between brain regions, the Dice overlap. Since the meta-analysis ALE map was produced by thresholding a map combining Gaussian kernels of various sizes \cite{gingerAle, gingerAle2}, we considered that the thresholded heatmap had to be smoothed, and we explored a broad range of thresholds and smoothings to search for the best possible match between meta-analysis and heatmaps. This kind of grid search is not common in the literature, but we think that it was justified to account for the unknown level of smoothing incorporated in the VBM studies and during their combination by GingeALE.

In addition to Dice overlaps, we computed Receiver Operating Characteristic (ROC) curves and precision-recall curves by considering the absolute values of the heatmaps as classification scores and the meta-analysis map as a set of binary labels. However, the regions affected by Alzheimer's disease cover only a very small part of the brain. As a result, the area under the ROC curves increased with heatmap smoothing to reach maxima corresponding to heatmaps selecting no voxels in the brain, and the ROC curves were of no use to select smoothing parameters. The second set of additional curves, the precision-recall curves, were not affected by this issue, but they selected similar optimal smoothing parameters to the grid search over Dice overlaps, as shown in the results presented in Supplementary materials (Section 4). Therefore, precision-recall curves provided few additional insights.

\section{Conclusion}
In this work, we evaluated the ability of three prominent CNN heatmap methods, the Layer-wise Relevance Propagation (LRP) method, the Integrated Gradients (IG) method, and the Guided Grad-CAM (GGC) method, to capture Alzheimer's disease effects in the ADNI data set by training CNN classifiers and measuring the overlap between their heatmaps and a brain map derived from a large-scale meta-analysis. We found that the three heatmap methods capture brain regions that overlap fairly well with the meta-analysis map, and we observed the best results for the IG method. All three heatmap methods outperformed linear SVM models. These results suggest that the analysis of deep nonlinear models by the most recent heatmap methods can produce more meaningful brain maps than linear and shallow models. Further work will be required to replicate our results and extend our models to investigate other tasks, such as other neurodegenerative disorders and healthy aging.

\section{Author Contributions}
D.W., N.H, and M.H. made substantial contributions to the analysis and interpretation of results. M.H. designed the study. D.W., N.H., and M.H drafted the article. All authors revised the article critically for important intellectual content, approved the version to be published, and agreed to be accountable for all aspects of the work in ensuring that questions related to the accuracy or integrity of any part of the work are appropriately investigated and resolved. 

\section{Acknowledgments}
This study was supported in part by the National Institute of Health (NIH) grant P30AG066546 (South Texas Alzheimer’s Disease Research Center) and grant numbers 5R01HL127659, 1U24AG074855, and the San Antonio Medical Foundation grant SAMF – 1000003860.

\section{Data Availability}
The brain scans used in the present work were obtained from the Alzheimer's Disease Neuroimaging Initiative (ADNI) database (adni.loni.usc.edu). The list of VBM studies combined to produce the meta-analysis map is provided in Supplementary materials (Section 2). The code is available at \href{https://github.com/UTHSCSA-NAL/CNN-heatmap}{https://github.com/UTHSCSA-NAL/CNN-heatmap/}.

\bibliographystyle{elsarticle-num}
\bibliography{biblio}
\end{document}

\title{Deep neural network heatmaps capture Alzheimer's disease patterns reported in a large meta-analysis of neuroimaging studies: Supplementary Material}

\author{Di Wang, Nicolas Honnorat, Peter T. Fox, Kerstin Ritter, \\ Simon B. Eickhoff, Sudha Seshadri, Mohamad Habes}

\date{}

\maketitle

\section{Dice overlaps for the synthetic data}
\begin{figure*}[!htb]
\centering
\includegraphics[width=1.0\linewidth]{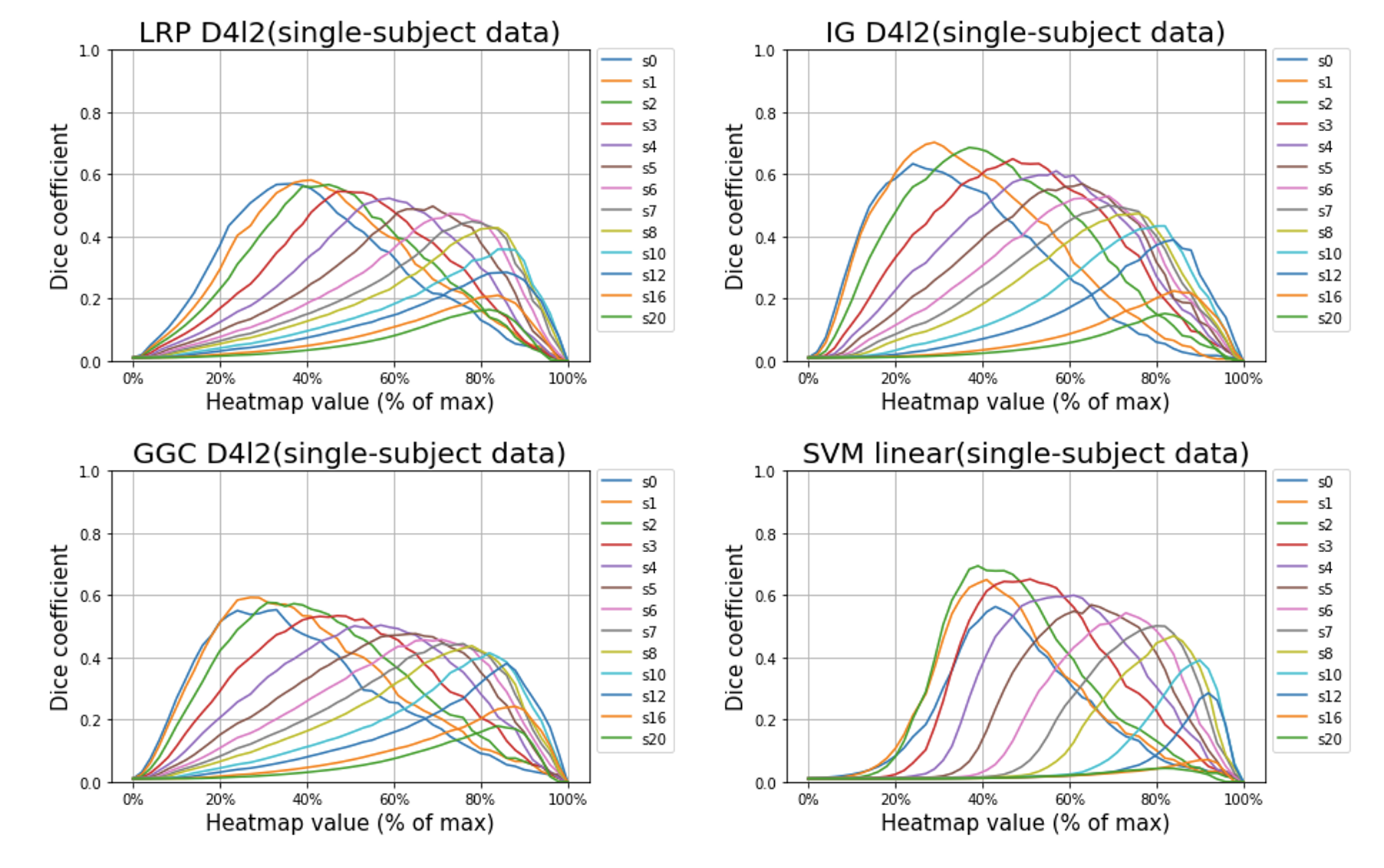}
\caption{\label{single-subject_dice}Dice overlap, for the single-subject simulated data, between all the smoothed heatmaps and the hippocampus map. s0 corresponds to the heatmaps before smoothing, s1 corresponds to the heatmaps after 1 mm FWHM Gaussian smoothing, and similarly for other smoothing FWHM.}
\end{figure*}

\begin{figure*}[!htb]
\includegraphics[width=1.0\linewidth]{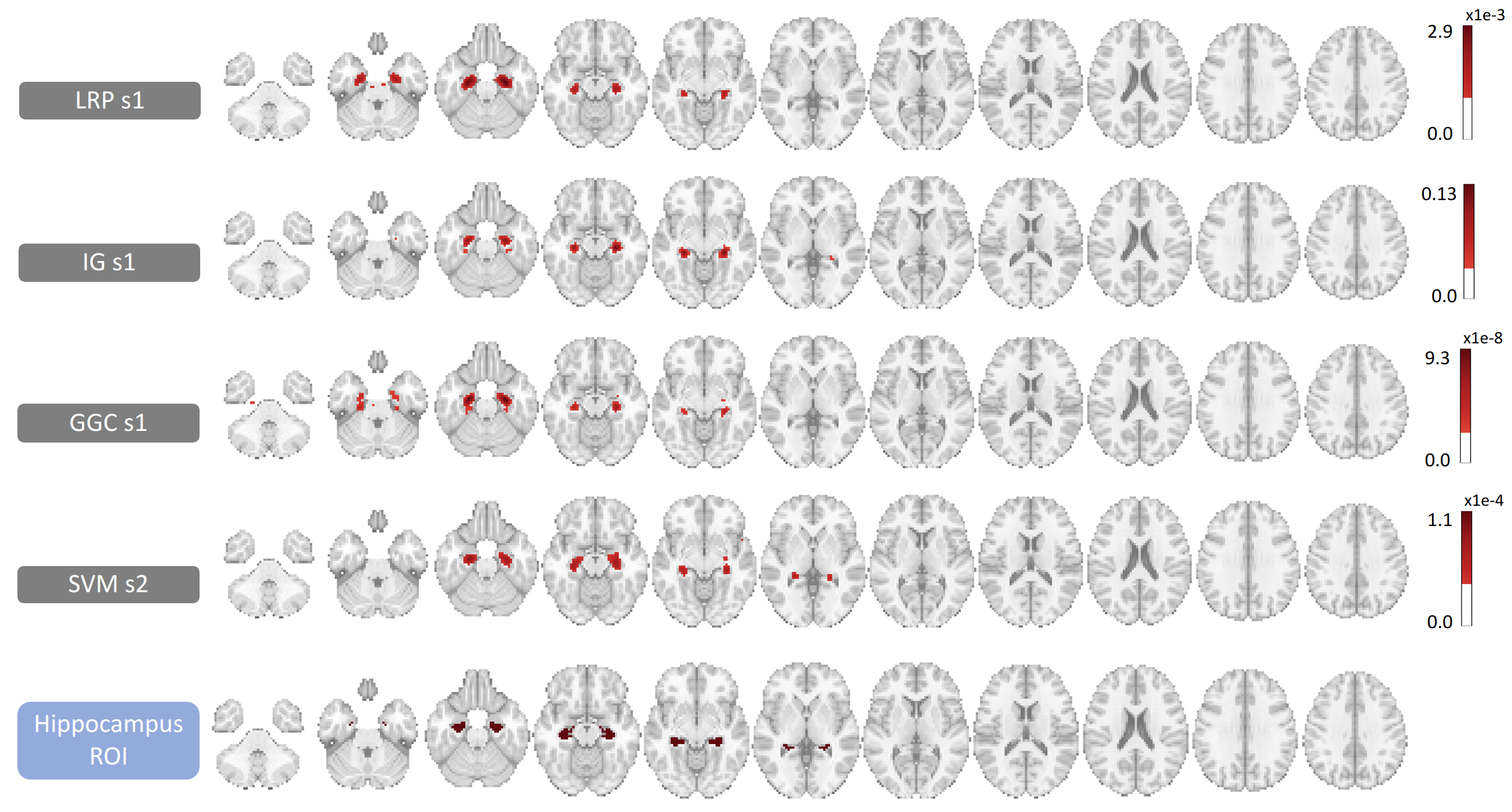}
\caption{\label{single-subject_bestdice}Heatmaps derived for the single-subject simulated data and corresponding to the best Dice overlap with the hippocampus region where the synthetic effect has been introduced, for CNN ModelD4l2 and linear SVM.}
\end{figure*}

\begin{figure*}[!htb]
\centering
\includegraphics[width=1.0\linewidth]{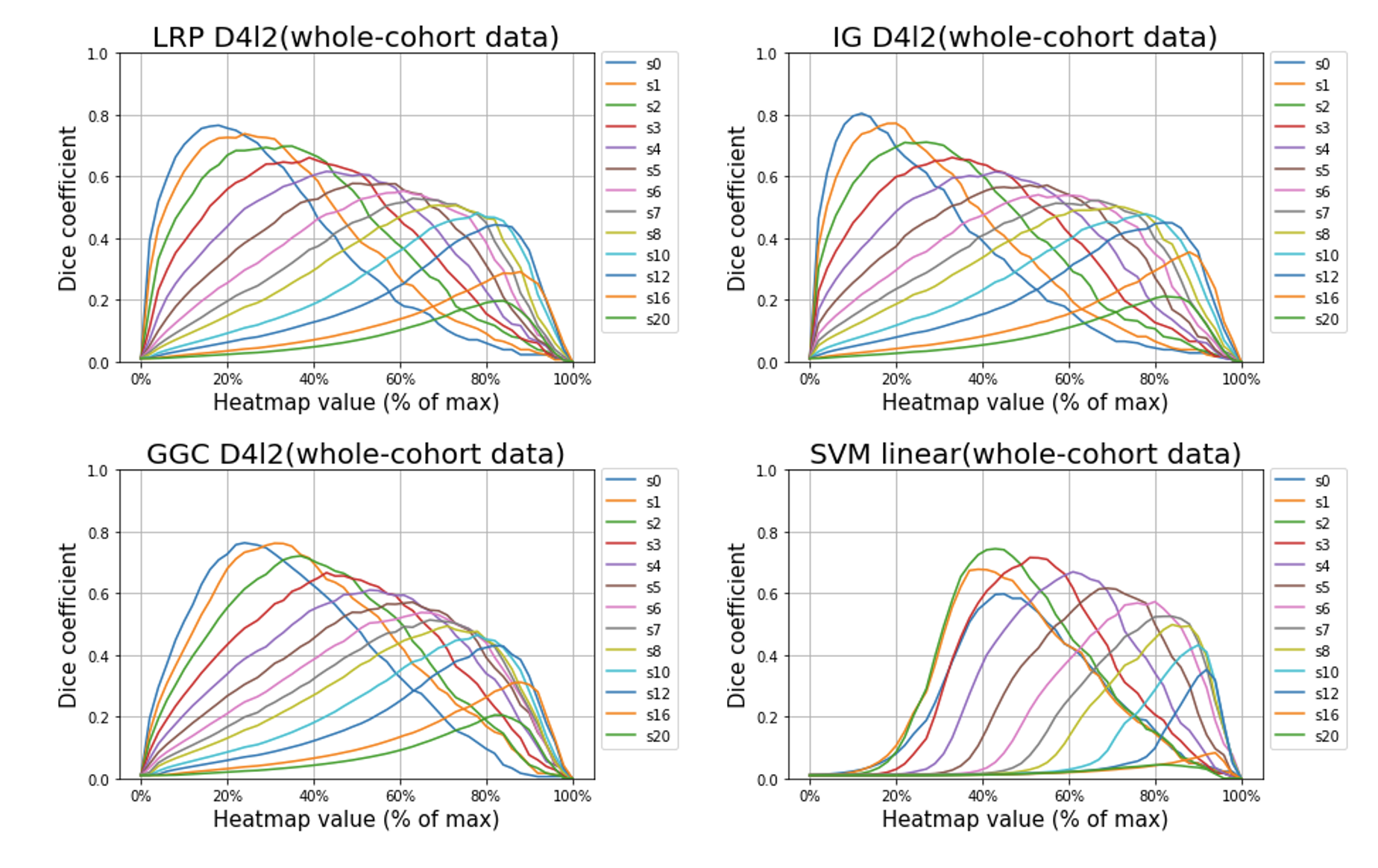}
\caption{\label{whole-cohort_dice}Dice overlap, for the whole-cohort simulated data, between all the smoothed heatmaps and the hippocampus map. s0 corresponds to the heatmaps before smoothing, s1 corresponds to the heatmaps after 1 mm FWHM Gaussian smoothing, and similarly for other smoothing FWHM.}
\end{figure*}

\begin{figure*}[!htb]
\includegraphics[width=1.0\linewidth]{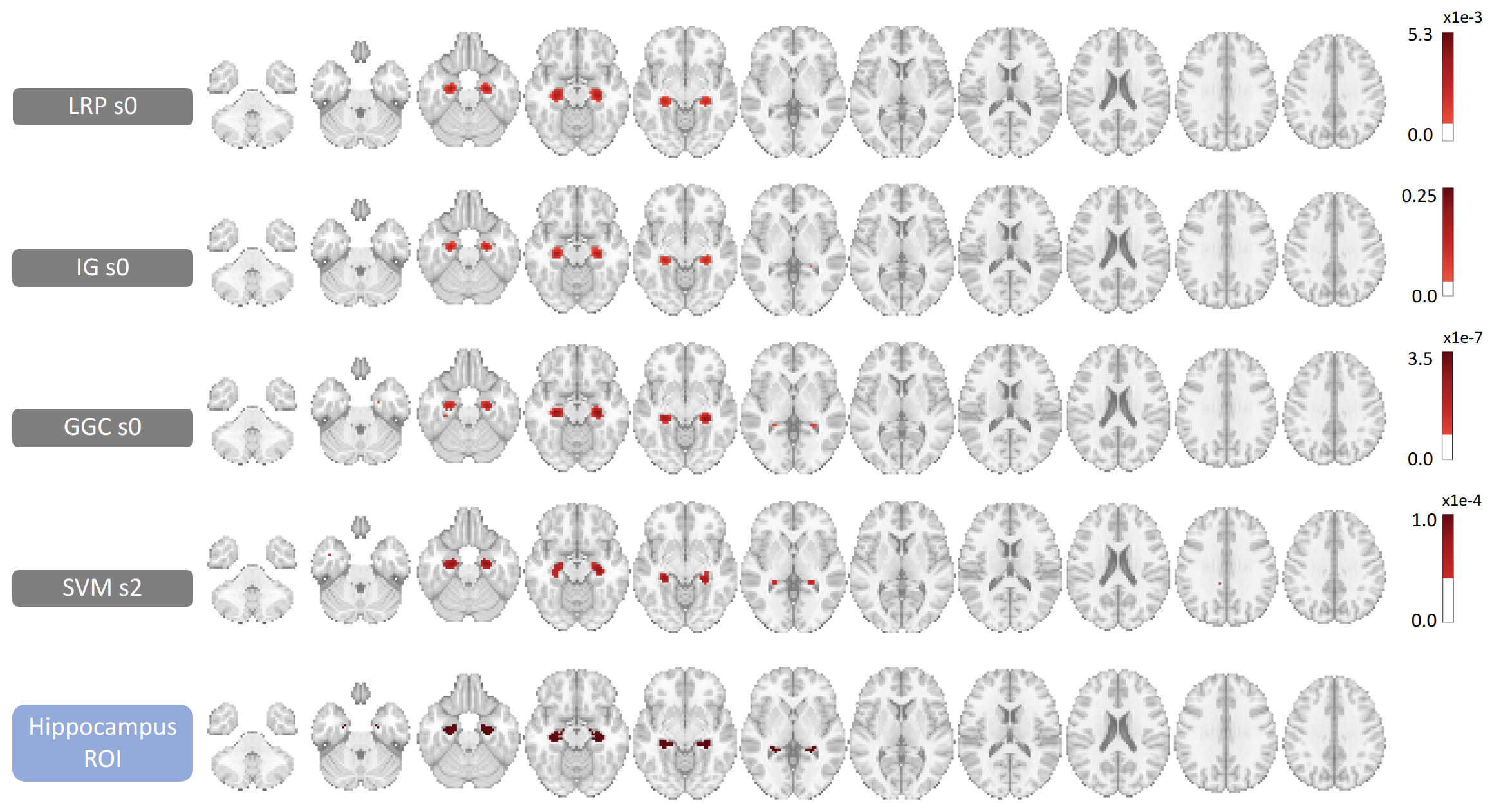}
\caption{\label{whole-cohort_bestdice}Heatmaps derived for the whole-cohort simulated data and corresponding to the best Dice overlap with the hippocampus region where the synthetic effect has been introduced, for CNN ModelD4l2 and linear SVM.}
\end{figure*}

\begin{figure*}[!htb]
\includegraphics[width=1.0\linewidth]{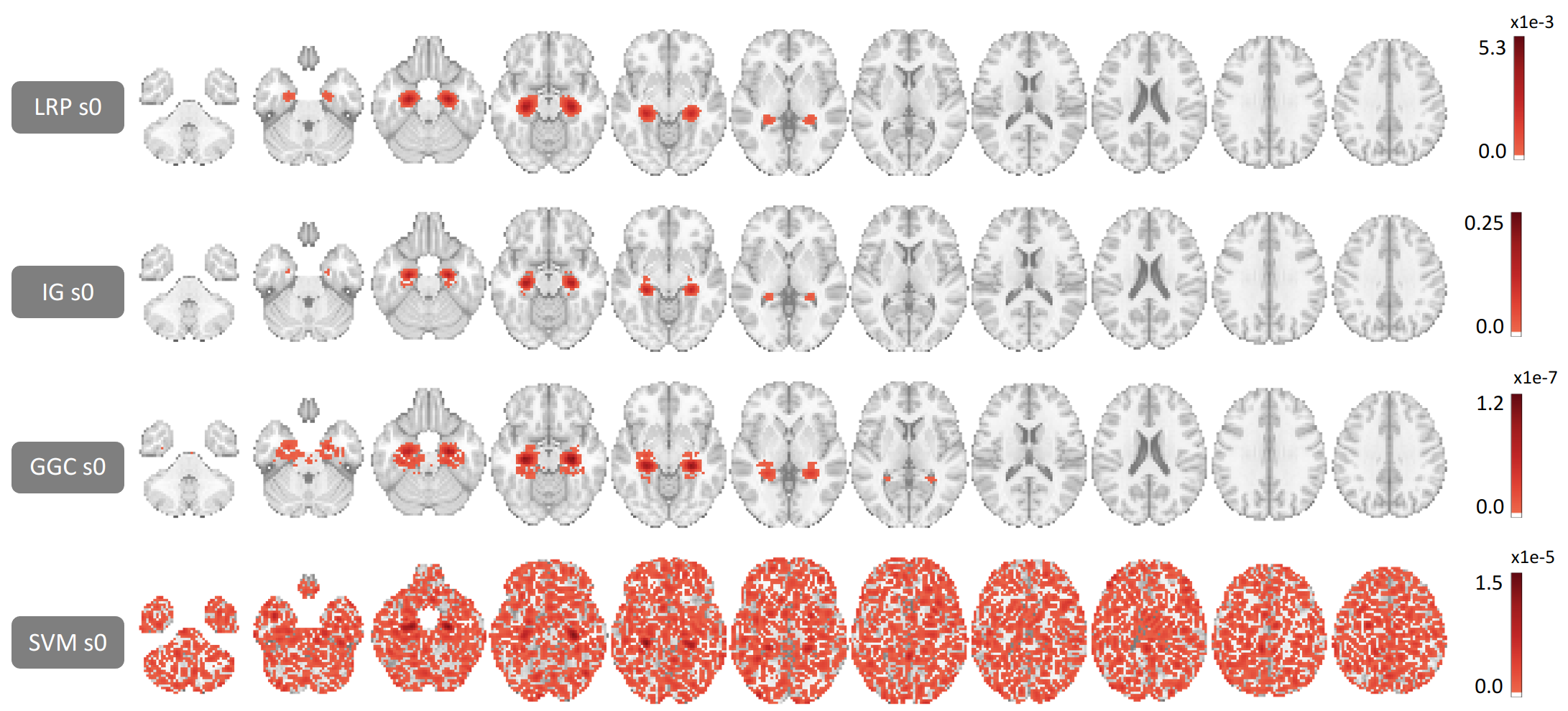}
\caption{\label{whole-cohort_5}Heatmaps derived for the whole-cohort simulated data, without smoothing, and thresholded at 5\% of their maximum value.}
\end{figure*}

The following Figures report the Dice overlaps between the hippocampus regions altered in the synthetic data sets and the brain regions captured by the coefficients of the best SVMs and the heatmap values of the best CNN classifiers, for the single-subject data set (Figure \ref{single-subject_dice}), and the whole-cohort data set (Figure \ref{whole-cohort_dice}). The heatmaps corresponding to the best overlaps are shown in Figure \ref{single-subject_bestdice} (for the single-subject data) and \ref{whole-cohort_bestdice} (for the whole-cohort data). Unsmoothed heatmaps thresholded at 5\% are shown in the following Figures (Figure \ref{single-subject_5} for single-subject, Figure \ref{whole-cohort_5} for whole-cohort).

\begin{figure*}[!htb]
\includegraphics[width=1.0\linewidth]{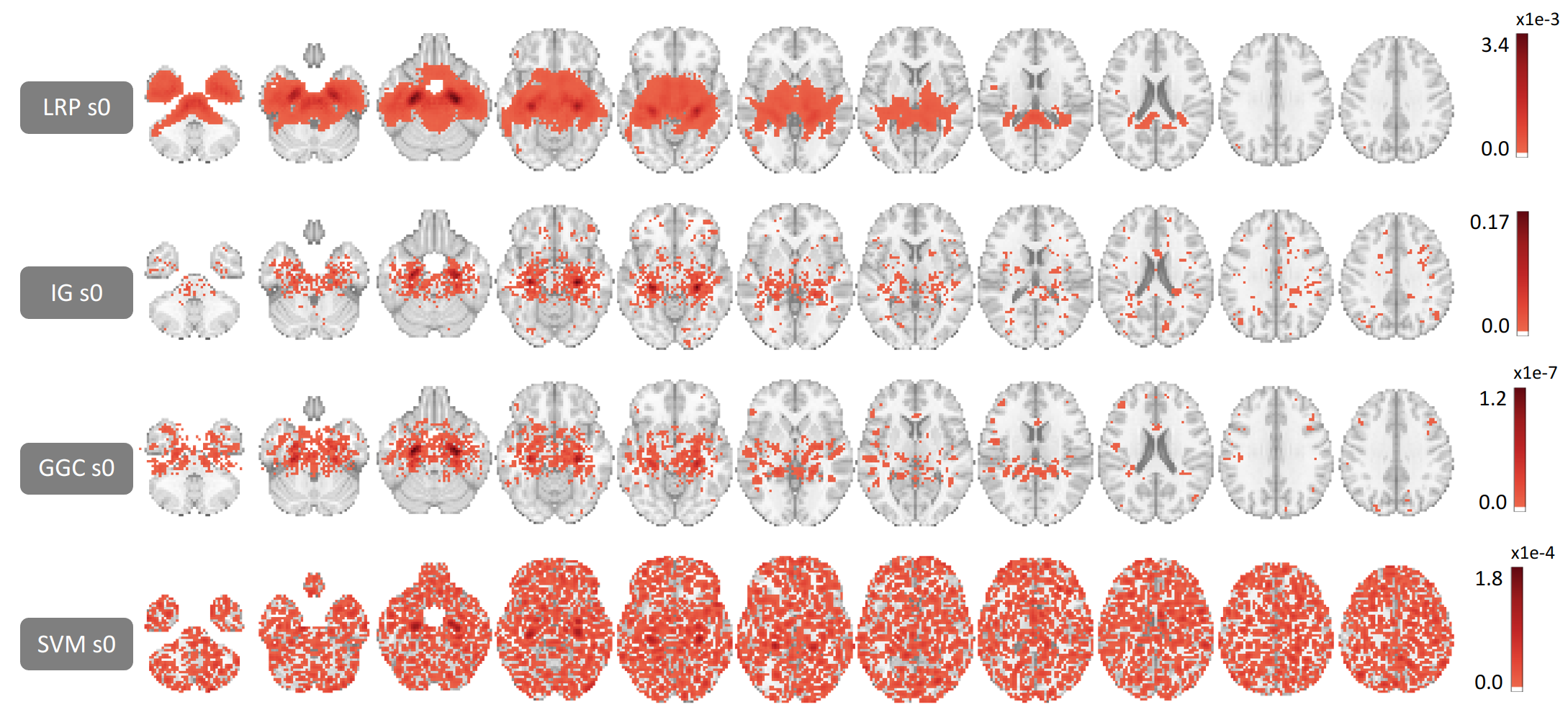}
\caption{\label{single-subject_5}Heatmaps derived for the single-subject simulated data, without smoothing, and thresholded at 5\% of their maximum value.}
\end{figure*}

\section{VBM studies included in the meta-analysis}
The voxel-based morphometry studies combined in our meta-analysis are reported in Table \ref{metainfo}. 

\captionsetup{width=12.0cm}
{\renewcommand{\arraystretch}{1.5}
\begin{longtable}[l]{|p{3.8cm}|p{1.9cm}|p{0.3cm}|p{4.2cm}|}
\hline
\textbf{Article} & \textbf{Experiment} & \textbf{$n$} & \textbf{Data source}\\ \hline
Barbeau E et al., 2008\textsuperscript{1}&MCI $<$ CN&28&Universit´e de Toulouse\\\hline
Baron JC et al., 2001\textsuperscript{2}&AD $<$ CN&19&University of Caen\\\hline
Baxter LC et al., 2006\textsuperscript{3}&AD $<$ CN&15&Sun Health Research Institute\\\hline
Bell-McGinty S et al., 2005\textsuperscript{4}&MCI $<$ CN&37&University of Pittsburgh\\\hline
Berlingeri M et al., 2008\textsuperscript{5}&AD $<$ CN&21&University of Milano-Bicocca\\\hline
Bozzali M et al., 2012\textsuperscript{6}&AD $<$ CN&31&IRCCS Fondazione Santa Lucia\\\hline
Bozzali M et al., 2006\textsuperscript{7}&Conv $<$ CN&14&IRCCS Fondazione Santa Lucia\\\hline
Bozzali M et al., 2006\textsuperscript{7}&Non $<$ CN&8&IRCCS Fondazione Santa Lucia\\\hline
Brambati et al., 2009\textsuperscript{8}&AD $<$ CN&10&McGill Center for Studies in Aging\\\hline
Brambati et al., 2009\textsuperscript{8}&MCI $<$ CN&11&McGill Center for Studies in Aging\\\hline
Brambati et al., 2009\textsuperscript{8}&MCI $<$ CN&14&McGill Center for Studies in Aging\\\hline
Brenneis C et al., 2004\textsuperscript{9}&AD $<$ CN&10&General Hospital of Linz\\\hline
Brys M et al., 2009\textsuperscript{10}&Conv $<$ CN&8&New York University School of Medicine, Center for Brain Health\\\hline
Canu E et al., 2012\textsuperscript{11}&EOAD $<$ CN&18&IRCCS Centro San Giovanni di Dio Fatebenefratelli\\\hline
Canu E et al., 2012\textsuperscript{11}&LOAD $<$ CN&24&IRCCS Centro San Giovanni di Dio Fatebenefratelli\\\hline
Caroli A et al., 2007\textsuperscript{12}&Conv $<$ CN&9&IRCCS San Giovanni di Dio-FBF\\\hline
Caroli A et al., 2007\textsuperscript{12}&Non $<$ CN&14&IRCCS San Giovanni di Dio-FBF\\\hline
Chen et al., 2011\textsuperscript{13}&AD $<$ CN&15&Pennsylvania’s Alzheimer’s Disease Center \\\hline
Chetelat G et al., 2002\textsuperscript{14}&AD $<$ CN&16&University of Caen\\\hline
Chetelat G et al., 2002\textsuperscript{14}&MCI $<$ CN&22&University of Caen\\\hline
Colloby SJ et al., 2014\textsuperscript{15}&AD $<$ CN&47&Newcastle University\\\hline
Defrancesco M et al., 2014\textsuperscript{16}&Conv $<$ CN&13&University Clinic of Innsbruck\\\hline
Derflinger S et al., 2011\textsuperscript{17}&AD $<$ CN&35&Technische Universitat Munchen, Munich, Germany\\\hline
Derflinger S et al., 2011\textsuperscript{17}&MCI $<$ CN&24&Technische Universitat Munchen, Munich, Germany\\\hline
Dos Santos V et al., 2011\textsuperscript{18}&AD $<$ CN&34&Heidelberg University\\\hline
Dos Santos V et al., 2011\textsuperscript{18}&MCI $<$ CN&60&Heidelberg University\\\hline
Farrow TFD et al., 2007\textsuperscript{19}&AD $<$ CN&7&North Sheffield Research\\\hline
Feldmann A et al., 2008\textsuperscript{20}&AD $<$ CN&6&local\\\hline
Ford AH et al., 2014\textsuperscript{21}&MCI $<$ CN&65&Perth Perception Study\\\hline
Gee J et al., 2003\textsuperscript{22}&AD $<$ CN&12&University of Pennsylvania Department of Neurology\\\hline
Gold BT et al., 2010\textsuperscript{23}&MCI $<$ CN&12&University of Kentucky\\\hline
Guo X et al., 2010\textsuperscript{24}&AD $<$ CN&13&Xuanwu Hospital\\\hline
Hamalainen A et al., 2007\textsuperscript{25}&MCI $<$ CN&43&University of Kuopio\\\hline
Hamalainen A et al., 2007\textsuperscript{25}&MCI $<$ CN&13&University of Kuopio\\\hline
Hamalainen A et al., 2007\textsuperscript{25}&AD $<$ CN&15&University of Kuopio\\\hline
Han Y et al., 2012\textsuperscript{26}&MCI $<$ CN&17&West China Hospital\\\hline
Hirao K et al., 2006\textsuperscript{27}&AD $<$ CN&61&National Center Hospital for Mental, Nervous and Muscular Disorders, National Center of Neurology and
Psychiatry, Tokyo, Japan\\\hline
Honea RA et al., 2009\textsuperscript{28}&AD $<$ CN&60&University of Kansas Brain Aging Project\\\hline
Hong YJ et al., 2015\textsuperscript{29}&MCI $<$ CN&29&Catholic University of Korea\\\hline
Hornberger et al., 2011\textsuperscript{30}&AD $<$ CN&15&FRONTIER database\\\hline
Huang CW et al., 2017\textsuperscript{31}&AD $<$ CN&50&Chang Gung Memorial Hospital\\\hline
Ibrahim I et al., 2009\textsuperscript{32}&AD $<$ CN&21&local\\\hline
Imabayashi E et al., 2013\textsuperscript{33}&AD $<$ CN&5&Japanese Alzheimer’s Disease Neuroimaging Initiative\\\hline
Kim S et al., 2011\textsuperscript{34}&AD $<$ CN&10&Chung-Ang University Hospita\\\hline
Kim S et al., 2011\textsuperscript{34}&AD $<$ CN&20&Chung-Ang University Hospita\\\hline
Kim S et al., 2011\textsuperscript{34}&AD $<$ CN&31&Chung-Ang University Hospita\\\hline
Lagarde J et al., 2015\textsuperscript{35}&AD $<$ CN&14&Salpetriere Hospital\\\hline
Mazere J et al., 2008\textsuperscript{36}&AD $<$ CN&8&University Hospital of Bordeaux\\\hline
Miettinen PS et al., 2011\textsuperscript{37}&AD $<$ CN&16&University of Eastern Finland\\\hline
Miettinen PS et al., 2011\textsuperscript{37}&MCI $<$ CN&18&University of Eastern Finland\\\hline
Migliaccio R et al., 2009\textsuperscript{38}&EOAD $<$ CN&16&University of California San Francisco\\\hline
Mitolo et al., 2013\textsuperscript{39}&MCI $<$ CN&20&University of Padua, Padua\\\hline
Mok et al., 2012\textsuperscript{40}&AD $<$ CN&14&Shin Kong Wu Ho-Su Memorial Hospita\\\hline
Mok et al., 2012\textsuperscript{40}&AD $<$ CN&10&Shin Kong Wu Ho-Su Memorial Hospita\\\hline
Pa et al., 2009\textsuperscript{41}&MCI $<$ CN&32&University of California San Francisco\\\hline
Pa et al., 2009\textsuperscript{41}&MCI $<$ CN&26&University of California San Francisco\\\hline
Pennanen C et al., 2005\textsuperscript{42}&MCI $<$ CN&51&Kuopio University Hospital\\\hline
Polat et al., 2012\textsuperscript{43}&AD $<$ CN&31&local\\\hline
Raji CA et al., 2009\textsuperscript{44}&AD $<$ CN&33&CHS\\\hline
Rami L et al., 2009\textsuperscript{45}&AD $<$ CN&31&local\\\hline
Rami L et al., 2009\textsuperscript{45}&MCI $<$ CN&14&local\\\hline
Remy F et al., 2005\textsuperscript{46}&AD $<$ CN&8&local\\\hline
Samuraki M et al., 2007\textsuperscript{47}&AD $<$ CN&39&Kanazawa University Hospital\\\hline
Saykin AJ et al., 2006\textsuperscript{48}&MCI $<$ CN&40&Dartmouth Medical School\\\hline
Schmidt-Wilcke T et al., 2009\textsuperscript{49}&MCI $<$ CN&18&local\\\hline
Shiino A et al., 2006\textsuperscript{50}&AD $<$ CN&40&Shiga University of Medical Science\\\hline
Shiino A et al., 2006\textsuperscript{50}&MCI $<$ CN&20&Shiga University of Medical Science\\\hline
Trivedi MA et al., 2006\textsuperscript{51}&MCI $<$ CN&15&University of Wisconsin\\\hline
Wang P et al., 2019\textsuperscript{52}&MCI $<$ CN&17&local\\\hline
Waragai M et al., 2009\textsuperscript{53}&AD $<$ CN&15&Tohoku University\\\hline
Whitwell JL et al., 2011\textsuperscript{54}&AD $<$ CN&14&Mayo Clinic\\\hline
Whitwell JL et al., 2011\textsuperscript{54}&AD $<$ CN&14&Mayo Clinic\\\hline
Xie S et al., 2006\textsuperscript{55}&AD $<$ CN&13&Peeking University First Hospital\\\hline
Yi D et al., 2015\textsuperscript{56}&MCI+ $<$ CN&10&Seoul National University\\\hline
Yi D et al., 2015\textsuperscript{56}&MCI- $<$ CN&10&Seoul National University\\\hline
Zahn R et al., 2005\textsuperscript{57}&AD $<$ CN&10&University of Freiburg\\\hline
Zhao Z et al., 2014\textsuperscript{58}&aMCI $<$ CN&20&Xuanwu Hospital\\\hline

\caption{List of publications included in the meta-analysis. AD: Alzheimer's disease, MCI: Mild cognitive impairment, Conv: MCI converters, Non: MCI non-converters, EOAD: Early-onset Alzheimer's disease, LOAD: Late-onset Alzheimer's disease, MCI+: MCI with high cerebral amyloid-beta protein (A$\beta$) deposition, MCI-: MCI with no or little cerebral amyloid-beta protein (A$\beta$) deposition. $n$ indicates the number of participants involved in each study.}\label{metainfo} \\
\end{longtable}
}

\section{Dice overlaps for the ADNI}
\begin{figure*}[!htb]
\centering
\includegraphics[width=1.0\linewidth]{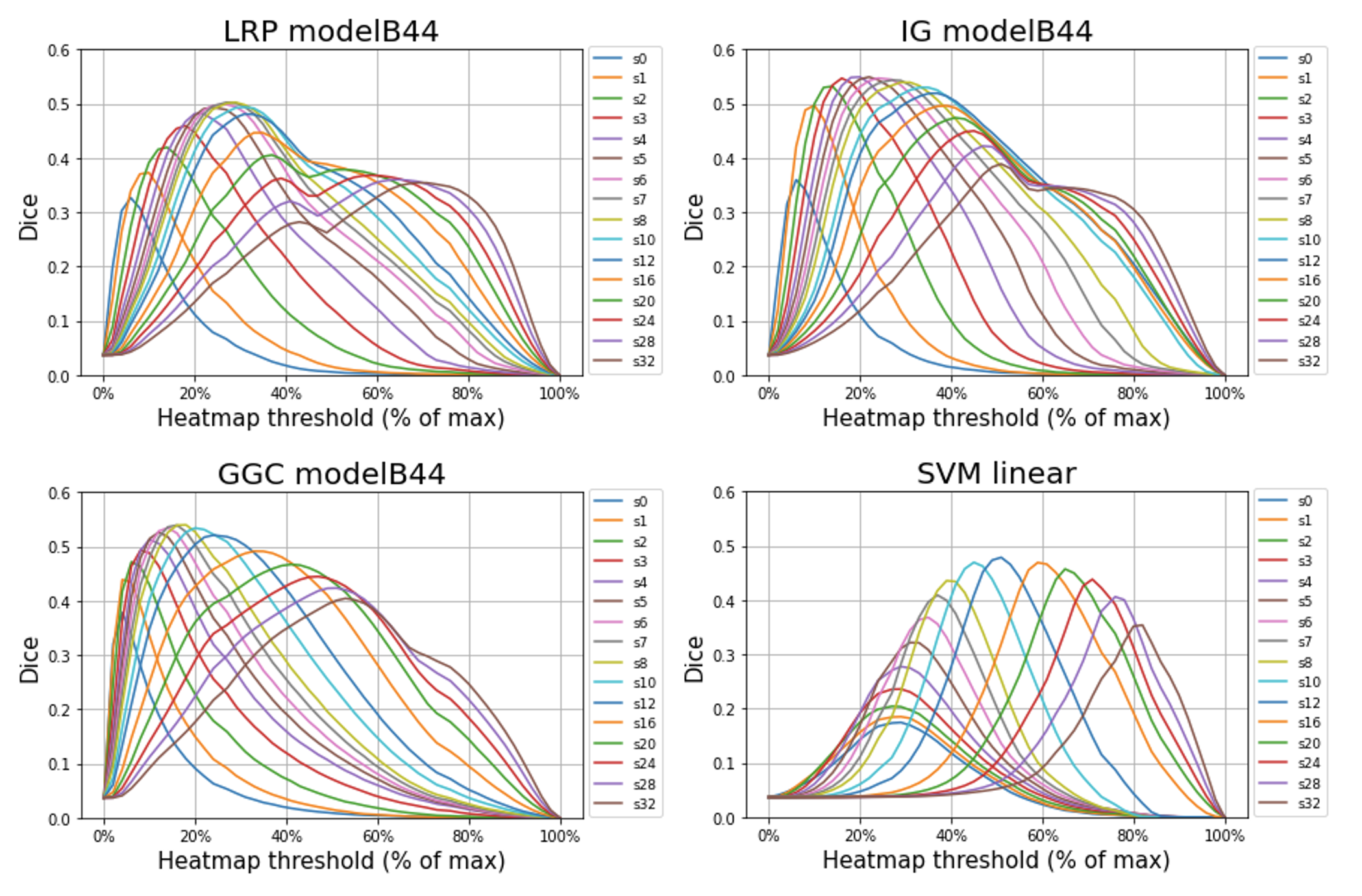}
\caption{\label{dice}Dice overlap between all the smoothed heatmaps and the meta-analysis map. s0 corresponds to the heatmaps before smoothing, s1 corresponds to the heatmaps after 1 mm FWHM Gaussian smoothing, and similarly for other smoothing FWHM.}
\end{figure*}

Figure \ref{dice} reports all the Dice overlap measured between the smoothed heatmaps derived for the modelB44 and the meta-analysis ALE map. ModelB44 achieved the best 5-fold cross-validation  of 87.24\% across all tested 3D CNN models.

\section{Additional heatmaps overlap measures for the ADNI}

In addition to the Dice overlaps shown in the previous section, the agreement between the meta-analysis brain map and the heatmaps extracted from the CNN model B44 was also measured with ROC curves and precision-recall curves as follows. For each voxel of each smoothed heatmap, the binary value of the meta-analysis map at that voxel was considered as a binary ground-truth label, and the value of the heatmap at that voxel as a score attempting to predict that label. A Receiver Operating Characteristic (ROC) curve and a precision-recall curve were then computed for each smoothed heatmap, and their area under the curve (AUC) was calculated. The overlaps obtained for the smoothed heatmaps were compared with the overlap measured when smoothing the SVM coefficients map.

\subsection{ModelB44 heatmaps ROC curves}
\begin{figure*}[h!]
\centering
\includegraphics[width=1.0\linewidth]{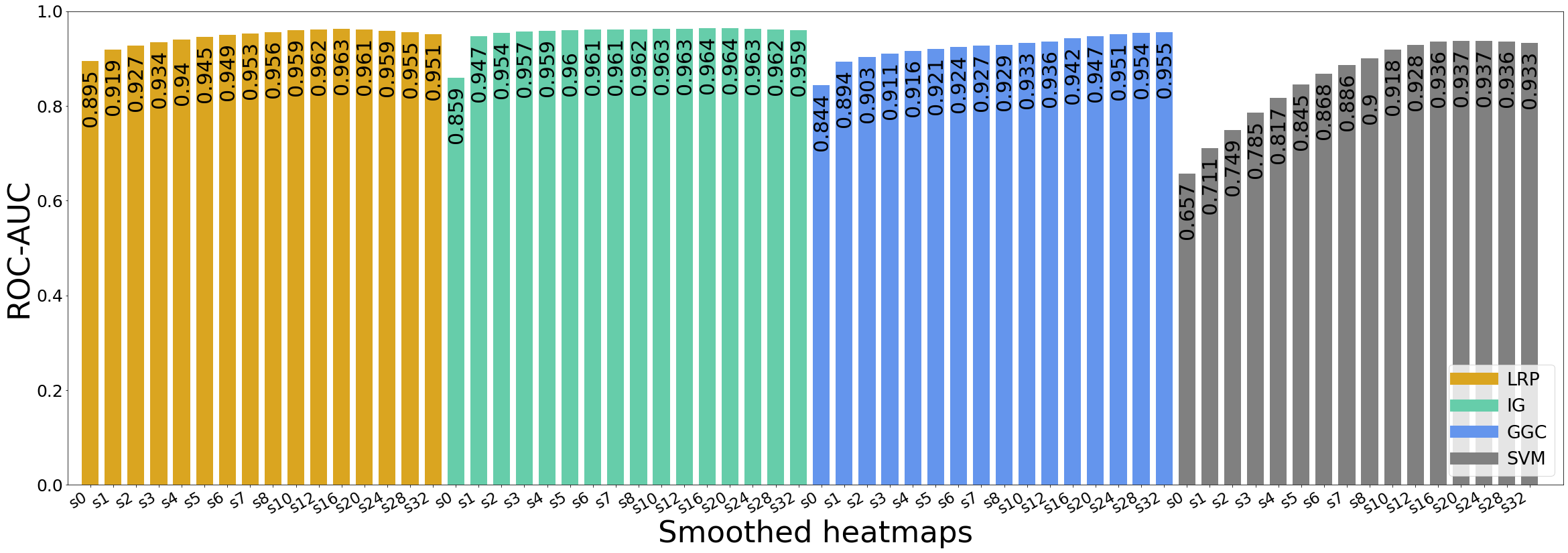} \\
\includegraphics[width=0.48\linewidth]{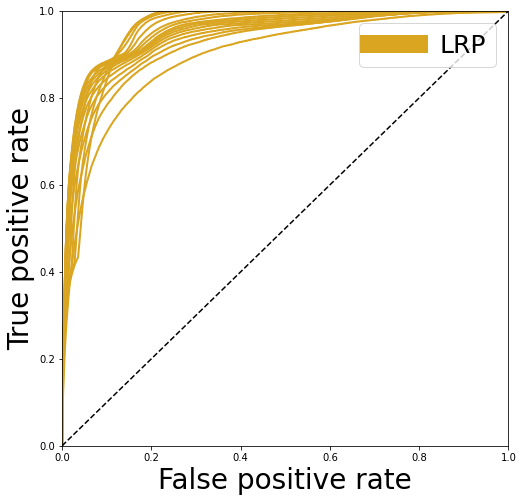}
\includegraphics[width=0.48\linewidth]{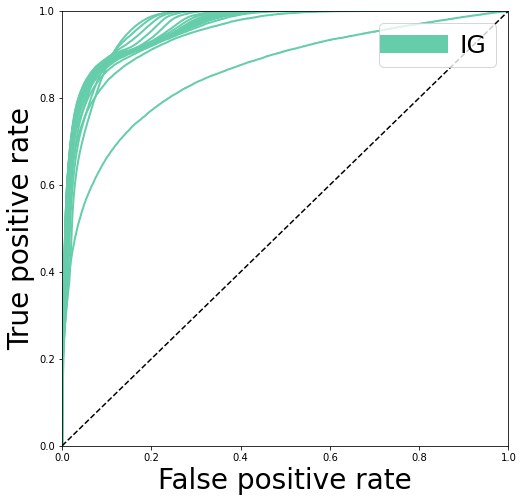} \\
\includegraphics[width=0.48\linewidth]{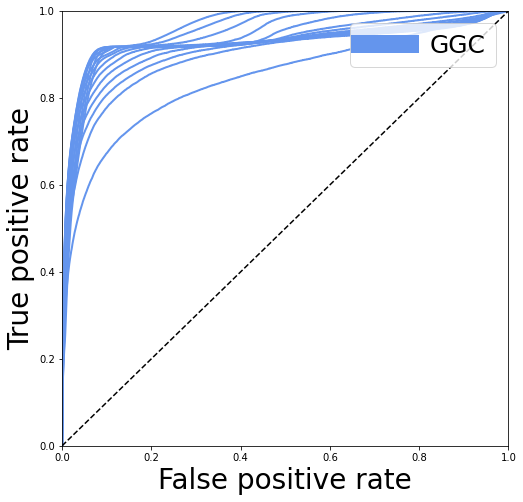}
\includegraphics[width=0.48\linewidth]{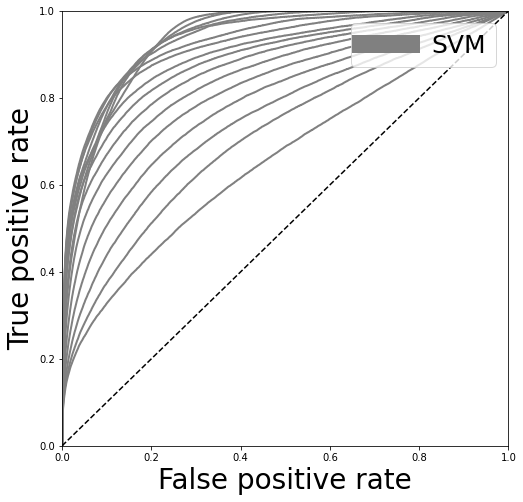}
\caption{\label{roc} ROC curves and their area under curves for all the heatmap methods and Gaussian smoothings.  All CNN heatmaps achieve better overlap with the meta-analysis map than the SVM heatmap.}
\end{figure*}

Figure \ref{roc} presents all the Receiver operating characteristic (ROC) curves obtained for the smoothed Model B44 heatmaps and the best SVM. All the methods reached better overlaps with the ground-truth meta-analysis map than the SVM maps. 

\subsection{ModelB44 heatmaps precision-recall curves}
\begin{figure*}[!htb]
\centering
\includegraphics[width=1.0\linewidth]{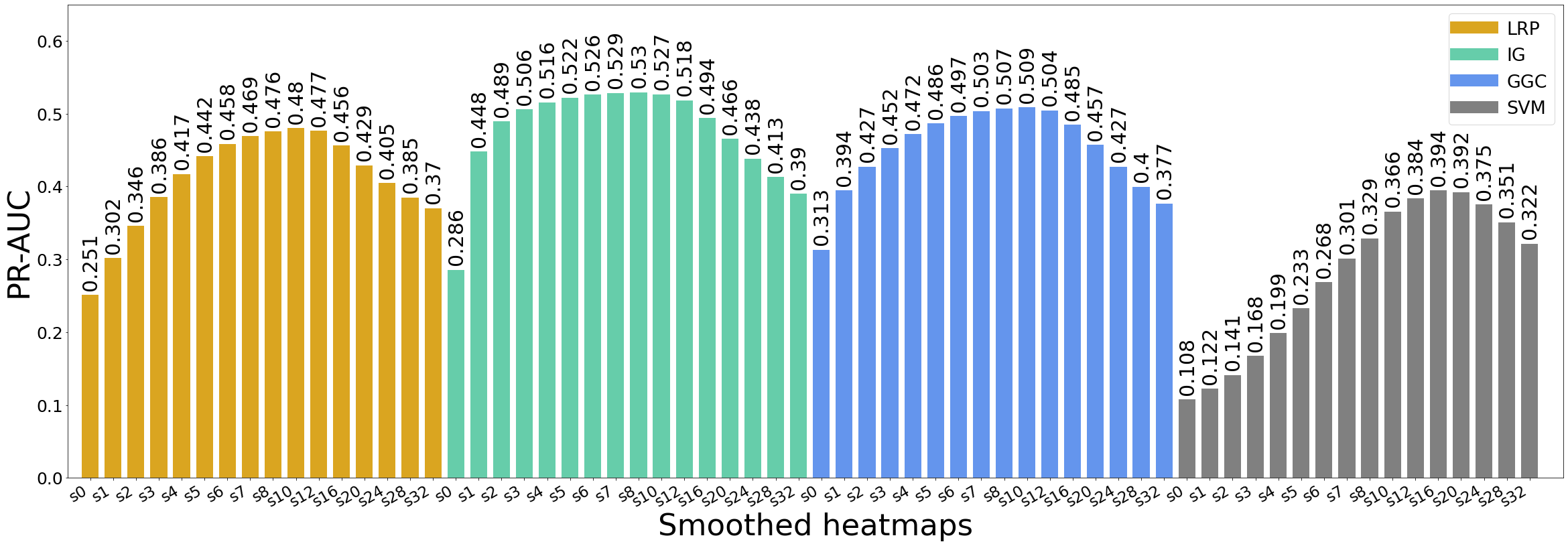} \\
\includegraphics[width=0.48\linewidth]{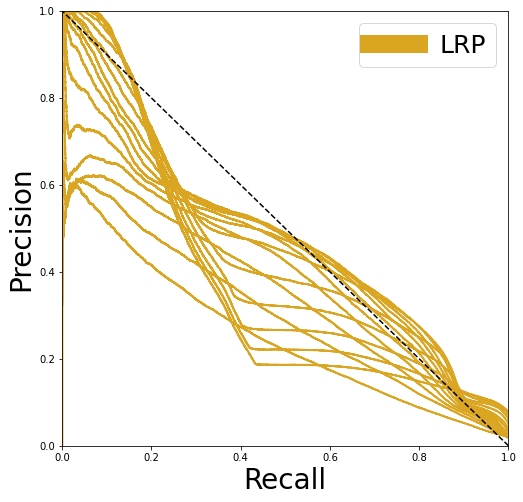}
\includegraphics[width=0.48\linewidth]{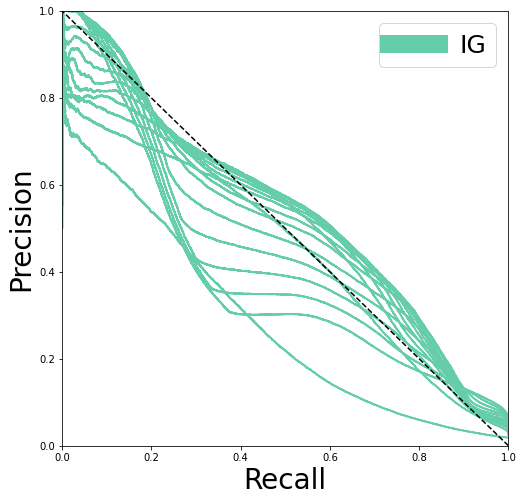} \\
\includegraphics[width=0.48\linewidth]{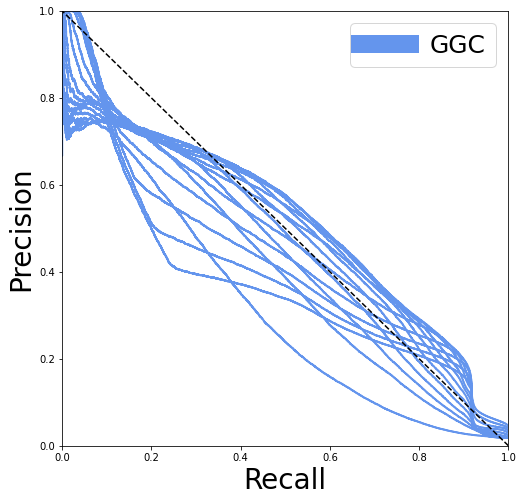}
\includegraphics[width=0.48\linewidth]{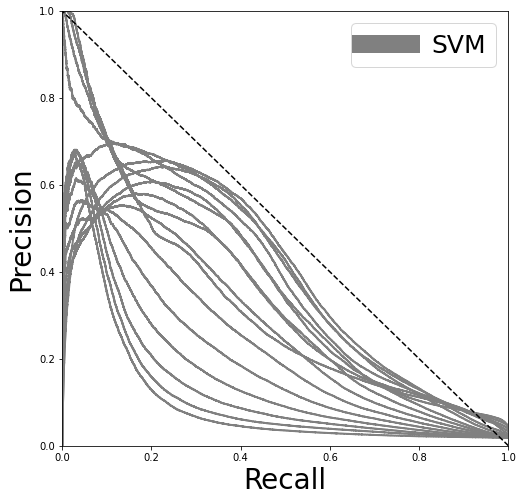}
\caption{\label{pr} Precision recall curves and their area under curves for all the heatmap methods and Gaussian smoothings. The best area under the curve is achieved for the GGC heatmap. LRP is the worst CNN heatmap. All CNN heatmaps achieve better overlap with the meta-analysis map than the SVM heatmap.}
\end{figure*}

Figure \ref{pr} presents all the precision-recall curves and precision-recall area-under-curve (PR-AUC) computed. This second set of results replicated the main findings of the previous Section: the three heatmap methods presented better overlap with the ground-truth meta-analysis map than the linear SVM map. The best PR-AUC was obtained for GGC, followed by IG. LRP is the heatmap method associated with the lowest overlap.   

The Gaussian smoothing producing the best Dice and PR-AUC overlaps were not the same, but they were selected within similar ranges and associated with similar Dice and PR-AUC measures for all the methods, as shown in the comparative Table \ref{pr_dice}. For instance, the SVM coefficient map reached its best Dice overlap with the meta-analysis map after being smoothed at FWHM 12 mm while the best PR-AUC was measured at 16 mm FWHM smoothing, but the Dice overlaps associated with these two smoothings were very similar: 0.479 and 0.469 respectively, as well as the PR-AUC (0.384 and 0.394). This observation suggests that the two criteria have selected similar and stable smoothing parameter values.

\begin{table*}[!htb]
\centering
\begin{tabular}{|c|c|c|}
\hline
Method & best Dice FWHM & best PR-AUC FWHM \\
\hline
LRP & ~\textbf{s7:} dice=0.502 pr-auc=0.469 & \textbf{s10:} dice=0.494 pr-auc=0.480\\
IG & ~\textbf{s4:} dice=0.550 pr-auc=0.516 & ~\textbf{s8:} dice=0.539 pr-auc=0.530\\
GGC & ~\textbf{s8:} dice=0.540 pr-auc=0.507 & ~\textbf{s10:} dice=0.534 pr-auc=0.509\\
SVM & \textbf{s12:} dice=0.479 pr-auc=0.384 & \textbf{s16:} dice=0.469 pr-auc=0.394\\
\hline
\end{tabular}
\caption{\label{pr_dice} Comparison between the Dice overlaps and PR-AUCs measured for the heatmap smoothings producing the overall best Dice overlap and the overall best PR-AUC, for each method separately. The Gaussian FWHM smoothing was in similar ranges and associated with performances often within a percent of each other.}
\end{table*}

\section{References}
\begin{enumerate}
\item
Barbeau EJ, Ranjeva JP, Didic M, Confort-Gouny S, Felician O, Soulier E, Cozzone PJ, Ceccaldi M, Poncet M. Profile of memory impairment and gray matter loss in amnestic mild cognitive impairment. Neuropsychologia. 2008;46(4):1009–1019.
\item
Baron JC, Chetelat G, Desgranges B, Perchey G, Landeau B, La Sayette V de, Eustache F. In vivo mapping of gray matter loss with voxel-based morphometry in mild Alzheimer's disease. NeuroImage. 2001; 14(2):298–309.
\item
Baxter LC, Sparks DL, Johnson SC, Lenoski B, Lopez JE, Connor DJ, Sabbagh MN. Relationship of cognitive measures and gray and white matter in Alzheimer's disease. J Alzheimers Dis. 2006;9(3):253–260.
% \item
% Beheshti I, Demirel H. Feature-ranking-based Alzheimer's disease classification from structural MRI. Magn Reson Imaging. 2016;34(3):252–263.
\item
Bell-McGinty S, Lopez OL, Meltzer CC, Scanlon JM, Whyte EM, Dekosky ST, Becker JT.  Differential cortical atrophy in subgroups of mild cognitive impairment. Arch Neurol. 2005; 62(9):1393–1397.
\item
Berlingeri M, Bottini G, Basilico S, Silani G, Zanardi G, Sberna M, Colombo N, Sterzi R, Scialfa G, Paulesu E. Anatomy of the episodic buffer: a voxel-based morphometry study in patients with dementia. Behav Neurol. 2008;19(1-2):29–34.
\item
Bozzali M, Giulietti G, Basile B, Serra L, Spano B, Perri R, Giubilei F, Marra C, Caltagirone C, Cercignani M. Damage to the cingulum contributes to Alzheimer's disease pathophysiology by deafferentation mechanism. Hum Brain Mapp. 2012;33(6):1295–1308.
\item
Bozzali M, Filippi M, Magnani G, Cercignani M, Franceschi M, Schiatti E, Castiglioni S, Mossini R, Falautano M, Scotti G, Comi G, Falini A. The contribution of voxel-based morphometry in staging patients with mild cognitive impairment. Neurol. 2006;67(3):453–460.
\item
Brambati SM, Belleville S, Kergoat M, Chayer C, Gauthier S, Joubert S. Single- and multiple-domain amnestic mild cognitive impairment: two sides of the same coin? Dement Geriatr Cogn Disord. 2009; 28(6):541–549.
\item
Brenneis C, Wenning GK, Egger KE, Schocke M, Trieb T, Seppi K, Marksteiner J, Ransmayr G, Benke T, Poewe W. Basal forebrain atrophy is a distinctive pattern in dementia with Lewy bodies. Neuroreport. 2004;15(11):1711–1714.
\item
Brys M, Glodzik L, Mosconi L, Switalski R, Santi S de, Pirraglia E, Rich K, Kim BC, Mehta P, Zinkowski R, Pratico D, Wallin A, Zetterberg H, Tsui WH, Rusinek H, Blennow K, Leon MJ de. Magnetic resonance imaging improves cerebrospinal fluid biomarkers in the early detection of Alzheimer's disease. J Alzheimers Dis. 2009;16(2):351–362.
\item
Canu E, Frisoni GB, Agosta F, Pievani M, Bonetti M, Filippi M. Early and late onset Alzheimer's disease patients have distinct patterns of white matter damage. Neurobiol Aging. 2012;33(6):1023–1033.
\item
Caroli A, Testa C, Geroldi C, Nobili F, Barnden LR, Guerra UP, Bonetti M, Frisoni GB. Cerebral perfusion correlates of conversion to Alzheimer's disease in amnestic mild cognitive impairment. J Neurol. 2007; 254(12): 1698–1707.
\item
Chen Y, Wolk DA, Reddin JS, et al. Voxel-level comparison of arterial spin-labeled perfusion MRI and FDG-PET in Alzheimer disease. Neurology. 2011;77(22):1977-1985.
\item
Chetelat G, Desgranges B, La Sayette V de, Viader F, Eustache F, Baron J. Mapping gray matter loss with voxel-based morphometry in mild cognitive impairment. Neuroreport. 2002;13(15):1939–1943.
\item
Colloby SJ, O'Brien JT, Taylor JP. Patterns of cerebellar volume loss in dementia with Lewy bodies and Alzheimers disease: A VBM-DARTEL study. Psychiatry Res. 2014;223(3):187-191.
\item
Defrancesco M, Egger K, Marksteiner J, Esterhammer R, Hinterhuber H, Deisenhammer EA, Schocke M. Changes in white matter integrity before conversion from mild cognitive impairment to Alzheimer's disease. PloS One. 2014;9(8):e106062.
\item
Derflinger S, Sorg C, Gaser C, Myers N, Arsic M, Kurz A, Zimmer C, Wohlschlager A, Muhlau M. Grey-matter atrophy in Alzheimer's disease is asymmetric but not lateralized. J Alzheimers Dis. 2011; 25(2):347–357.
\item
Dos Santos V, Thomann PA, Wustenberg T, Seidl U, Essig M, Schroder J. Morphological cerebral correlates of CERAD test performance in mild cognitive impairment and Alzheimer's disease. J Alzheimers Dis. 2011;23(3):411–420.
\item
Farrow TFD, Thiyagesh SN, Wilkinson ID, Parks RW, Ingram L, Woodruff PWR. Fronto-temporal-lobe atrophy in early-stage Alzheimer's disease identified using an improved detection methodology. Psychiatry Res. 2007; 155(1):11–19.
\item
Feldmann A, Trauninger A, Toth L, et al. Atrophy and decreased activation of fronto-parietal attention areas contribute to higher visual dysfunction in posterior cortical atrophy. Psychiatry Res. 2008; 164(2):178-184.
\item
Ford AH, Almeida OP, Flicker L, Garrido GJ, Greenop KR, Foster JK, Etherton-Beer C, van Bockxmeer FM, Lautenschlager NT. Grey matter changes associated with deficit awareness in mild cognitive impairment: a voxel-based morphometry study. J Alzheimers Dis. 2014; 42(4):1251–1259.
\item
Gee J, Ding L, Xie Z, Lin M, DeVita C, Grossman M. Alzheimer's disease and frontotemporal dementia exhibit distinct atrophy-behavior correlates: a computer-assisted imaging study. Acad Radiology. 2003; 10(12):1392–1401.
\item
Gold BT, Jiang Y, Jicha GA, Smith CD. Functional response in ventral temporal cortex differentiates mild cognitive impairment from normal aging. Hum Brain Mapp. 2010;31(8):1249–1259.
\item
Guo X, Wang Z, Li K, Li Z, Qi Z, Jin Z, Yao L, Chen K. Voxel-based assessment of gray and white matter volumes in Alzheimer's disease. Neurosci Lett. 2010;468(2):146–150.
\item
Hamalainen A, Tervo S, Grau-Olivares M, et al. Voxel-based morphometry to detect brain atrophy in progressive mild cognitive impairment. Neuroimage. 2007;37(4):1122-1131.
\item
Han Y, Lui S, Kuang W, Lang Q, Zou L, Jia J. Anatomical and functional deficits in patients with amnestic mild cognitive impairment. PloS One. 2012;7(2):e28664.
\item
Hirao K, Ohnishi T, Matsuda H, Nemoto K, Hirata Y, Yamashita F, Asada T, Iwamoto T. Functional interactions between entorhinal cortex and posterior cingulate cortex at the very early stage of Alzheimer's disease using brain perfusion single-photon emission computed tomography. Nucl Med Commun. 2006;27(2):151–156.
\item
Honea RA, Thomas GP, Harsha A, Anderson HS, Donnelly JE, Brooks WM, Burns JM. Cardiorespiratory fitness and preserved medial temporal lobe volume in Alzheimer disease. Alzheimer Dis Assoc Disord. 2009;23(3):188–197.
\item
Hong YJ, Yoon B, Shim YS, Ahn KJ, Yang DW, Lee J. Gray and White Matter Degenerations in Subjective Memory Impairment: Comparisons with Normal Controls and Mild Cognitive Impairment. J Kor Med Sci. 2015;30(11):1652–1658.
\item
Hornberger M, Geng J, Hodges, JR. Convergent grey and white matter evidence of orbitofrontal cortex changes related to disinhibition in behavioural variant frontotemporal dementia. Brain. 2011; 134(Pt 9): 2502–2512.
\item
Huang CW, Hsu SW, Chang YT, et al. Cerebral Perfusion Insufficiency and Relationships with Cognitive Deficits in Alzheimer's Disease: A Multiparametric Neuroimaging Study. Sci Rep. 2018;8(1):1541.
\item
Ibrahim I, Horacek J, Bartos A, Hajek M, Ripova D, Brunovsky M, Tintera J. Combination of voxel based morphometry and diffusion tensor imaging in patients with Alzheimer's disease. Neurol Endocrin Lett. 2009;30(1):39–45.
\item
Imabayashi E, Matsuda H, Tabira T, et al. Comparison between brain CT and MRI for voxel-based morphometry of Alzheimer's disease. Brain Behav. 2013;3(4):487-493.
\item
Kim S, Youn YC, Hsiung GY, Ha SY, Park KY, Shin HW, Kim DK, Kim SS, Kee BS. Voxel-based morphometric study of brain volume changes in patients with Alzheimer's disease assessed according to the Clinical Dementia Rating score. J Clin Neurosci. 2011;18(7):916–921.
\item
Lagarde J, Valabregue R, Corvol JC, Garcin B, Volle E, Le Ber I, Vidailhet M, Dubois B, Levy R. Why do patients with neurodegenerative frontal syndrome fail to answer: 'In what way are an orange and a banana alike?'. Brain. 2015;138(Pt 2):456–471.
\item
Mazere J, Prunier C, Barret O, et al. In vivo SPECT imaging of vesicular acetylcholine transporter using [(123)I]-IBVM in early Alzheimer's disease. Neuroimage. 2008;40(1):280-288.
\item
Miettinen PS, Pihlajamaki M, Jauhiainen AM, et al. Structure and function of medial temporal and posteromedial cortices in early Alzheimer's disease. Eur J Neurosci. 2011;34(2):320-330.
\item
Migliaccio R, Agosta F, Possin KL, Canu E, Filippi M, Rabinovici GD, Rosen HJ, Miller BL, Gorno-Tempini ML. Mapping the Progression of Atrophy in Early- and Late-Onset Alzheimer's Disease. J Alzheimers Dis. 2015;46(2):351–364.
\item
Mitolo M, Gardini S, Fasano F, Crisi G, Pelosi A, Pazzaglia F, Caffarra P. Visuospatial memory and neuroimaging correlates in mild cognitive impairment. J Alzheimers Dis. 2013;35(1):75–90.
\item
Mok GS, Wu YY, Lu KM, Wu J, Chen LK, Wu TH. Evaluation of the screening power of Cognitive Abilities Screening Instrument for probable Alzheimer's disease using voxel-based morphometry. Clin Imaging. 2012;36(1):46-53.
\item
Pa J, Boxer A, Chao LL, Gazzaley A, Freeman K, Kramer J, Miller BL, Weiner MW, Neuhaus J, Johnson JK. Clinical-neuroimaging characteristics of dysexecutive mild cognitive impairment. Ann Neurol. 2009; 65(4): 414–423.
\item
Pennanen C, Testa C, Laakso MP, Hallikainen M, Helkala E, Hanninen T, Kivipelto M, Kononen M, Nissinen A, Tervo S, Vanhanen M, Vanninen R, Frisoni GB, Soininen H. A voxel based morphometry study on mild cognitive impairment. J Neurol Neurosurg Psychiatry. 2005;76(1):11–14.
\item
Polat F, Demirel SO, Kitis O, Simsek F, Haznedaroglu DI, Coburn K, Kumral E, Gonul AS. Computer based classification of MR scans in first time applicant Alzheimer patients. Curr Alzheimer Res. 2012; 9(7):789–794.
\item
Raji CA, Lopez OL, Kuller LH, Carmichael OT, Becker JT. Age, Alzheimer disease, and brain structure. Neurology. 2009; 73(22): 1899-1905.
\item
Rami L, Gomez-Anson B, Monte GC, Bosch B, Sanchez-Valle R, Molinuevo JL. Voxel based morphometry features and follow-up of amnestic patients at high risk for Alzheimer's disease conversion. Int J Geriatr Psychiatry. 2009;24(8):875-884.
\item
Rémy F, Mirrashed F, Campbell B, Richter W. Verbal episodic memory impairment in Alzheimer's disease: a combined structural and functional MRI study. NeuroImage. 2005;25(1):253–266.
\item
Samuraki M, Matsunari I, Chen WP, Yajima K, Yanase D, Fujikawa A, Takeda N, Nishimura S, Matsuda H, Yamada M. Partial volume effect-corrected FDG PET and grey matter volume loss in patients with mild Alzheimer's disease. Eur J Nucl Med Mol Imaging. 2007; 34(10): 1658–1669.
\item
Saykin AJ, Wishart HA, La Rabin, Santulli RB, La Flashman, West JD, McHugh TL, Mamourian AC. Older adults with cognitive complaints show brain atrophy similar to that of amnestic MCI. Neurol. 2006;67(5):834–842.
\item
Schmidt-Wilcke T, Poljansky S, Hierlmeier S, Hausner J, Ibach B. Memory performance correlates with gray matter density in the ento-/perirhinal cortex and posterior hippocampus in patients with mild cognitive impairment and healthy controls a voxel based morphometry study. NeuroImage. 2009; 47(4): 1914–1920.
\item
Shiino A, Watanabe T, Maeda K, Kotani E, Akiguchi I, Matsuda M. Four subgroups of Alzheimer's disease based on patterns of atrophy using VBM and a unique pattern for early onset disease. NeuroImage. 2006;33(1):17–26.
\item
Trivedi MA, Wichmann AK, Torgerson BM, Ward MA, Schmitz TW, Ries ML, Koscik RL, Asthana S, Johnson SC. Structural MRI discriminates individuals with Mild Cognitive Impairment from age-matched controls: a combined neuropsychological and voxel based morphometry study. Alzheimers Dement. 2006;2(4):296–302.
\item
Wang P, Li R, Yu J, Huang Z, Li J. Frequency-Dependent Brain Regional Homogeneity Alterations in Patients with Mild Cognitive Impairment during Working Memory State Relative to Resting State. Front Aging Neurosci. 2016 Mar 24;8:60. 
\item
Waragai M, Okamura N, Furukawa K, Tashiro M, Furumoto S, Funaki Y, Kato M, Iwata R, Yanai K, Kudo Y, Arai H. Comparison study of amyloid PET and voxel-based morphometry analysis in mild cognitive impairment and Alzheimer's disease. J Neurol Sci. 2009;285(1-2):100–108.
\item
Whitwell JL, Jack CR, JR, Przybelski SA, Parisi JE, Senjem ML, Boeve BF, Knopman DS, Petersen RC, Dickson DW, Josephs KA. Temporoparietal atrophy: a marker of AD pathology independent of clinical diagnosis. Neurobiol Aging. 2011; 32(9): 1531–1541.
\item
Xie S, Xiao JX, Gong GL, Zang YF, Wang YH, Wu HK, Jiang XX. Voxel-based detection of white matter abnormalities in mild Alzheimer disease. Neurol. 2006;66(12):1845–1849.
\item
Yi D, Choe YM, Byun MS, Sohn BK, Seo EH, Han J, Park J, Woo JI, Lee DY. Differences in functional brain connectivity alterations associated with cerebral amyloid deposition in amnestic mild cognitive impairment. Front Aging Neurosci. 2015;7:15.
\item
Zahn R, Buechert M, Overmans J, Talazko J, Specht K, Ko CW, Thiel T, Kaufmann R, Dykierek P, Juengling F, Hull M. Mapping of temporal and parietal cortex in progressive nonfluent aphasia and Alzheimer's disease using chemical shift imaging, voxel-based morphometry and positron emission tomography. Psychiatry Res. 2005;140(2):115–131.
\item
Zhao Z, Lu J, Jia X, Chao W, Han Y, Jia J, Li K. Selective changes of resting-state brain oscillations in aMCI: an fMRI study using ALFF. Biomed Res Int. 2014;2014:920902. 
\end{enumerate}